\documentclass[reprint, secnumarabic, amssymb, nobibnotes, aps, pra]{revtex4-1}
\usepackage[utf8]{inputenc}

 \usepackage{amsmath}
 \usepackage{bm}
 \usepackage{array}
 \usepackage{calc}
 \usepackage[dvipsnames]{xcolor}
 \usepackage{hyperref}
 \usepackage{ifthen}
 \usepackage{gensymb}
 \usepackage{graphicx}
 \usepackage{float}
 \usepackage{tikz}
 \usetikzlibrary{angles,calc,matrix,positioning,through,arrows.meta}
 \usepackage{tikz-layers}
 
\DeclareMathOperator{\Arg}{Arg}

\DeclareMathOperator{\sgn}{sign}

\newcommand{\bfhat}[1]{\boldsymbol{\hat{#1}}}

\newcommand{\odl}{(\Omega_{xy} \hat{L}_{xy} + \Omega_{zw} \hat{L}_{zw})}
\newcommand{\diff}[1]{\mathrm{d}#1}
\newcommand{\ta}{\theta}

\newcommand{\dint}{\int\displaylimits}

\newcommand{\cint}[1][0]{
    \ifthenelse{\equal{#1}{0}}{
        \dint_0^{2\pi}
    }{
        \dint_{#1}^{#1+2\pi}
    }
}
\newcommand{\cintm}[1][0]{
    \ifthenelse{\equal{#1}{0}}{
        \dint_{-\pi}^{\pi}
    }{
        \dint_{#1-\pi}^{#1+\pi}
    }
}

\newcommand{\abs}[1]{\lvert#1\rvert}

\newcommand{\acr}{\acute{r}}
\newcommand{\act}{\acute{\ta}}
\newcommand{\grr}{\grave{r}}
\newcommand{\grt}{\grave{\ta}}

\newcommand{\cpolar}[2][]{r_{#2}#1e^{i\ta_{#2}#1}}
\newcommand{\ccpolar}[2][]{r_{#2}#1e^{-i\ta_{#2}#1}}

\newcommand{\en}{\text{--}}

\newcommand{\acbf}[1]{\acute{\mathbf{#1}}}
\newcommand{\grbf}[1]{\grave{\mathbf{#1}}}
\newcommand{\vDimens}{\frac{\hbar}{m}}

\begin{document}

\title{Curved vortex surfaces in four-dimensional superfluids: \\
II. Equal-frequency double rotations}
\author{Ben McCanna}
	\email{benmccanna@outlook.com}
	\affiliation{School of Physics and Astronomy, University of Birmingham, Edgbaston Park Road, B15 2TT, West Midlands, United Kingdom}
	\author{Hannah M. Price}
			\email{H.Price.2@bham.ac.uk}
	\affiliation{School of Physics and Astronomy, University of Birmingham, Edgbaston Park Road, B15 2TT, West Midlands, United Kingdom}

\begin{abstract}
As is well-known, two-dimensional and three-dimensional superfluids under rotation can support topological excitations such as quantized point vortices and line vortices respectively. Recently, we have studied how, in a hypothetical four-dimensional (4D) superfluid, such excitations can be generalised to vortex planes and surfaces. In this paper, we continue our analysis of skewed and curved vortex surfaces based on the 4D Gross-Pitaevskii equation, and show that certain types of such states can be stabilised by equal-frequency double rotations for suitable parameters. This work extends the rich phenomenology of vortex surfaces in 4D, and raises interesting questions about vortex reconnections and the competition between various vortex structures which have no direct analogue in lower dimensions. 
	\end{abstract}
\maketitle

\section{Introduction}

Quantum vortices are topological excitations of a superfluid that are characterised by a quantized circulation around the  ``vortex core", where the superfluid density vanishes~\cite{pitaevskii2003, pethick2002, cooper2008rapidly, fetter2009, madison2000, madison2001, matthews1999,abo2001observation,verhelst2017vortex}. In 2D, this vortex core can be thought of as an effectively zero-dimensional point, while in 3D, the core becomes a one-dimensional line. In a hypothetical 4D superfluid, such a core can then extend into a two-dimensional plane or surface, which can have a much more varied phenomenology~\cite{mccanna2021, mccannaunequal}.  

As vortices are excitations, they are associated with an energy cost, which can be offset, for example, through rotation of the superfluid~\cite{fetter2009,cooper2008rapidly}, or equivalently by applying artificial magnetic fields~\cite{dalibard2011colloquium,Cooper_2019,Ozawa2019Photonics,price2022roadmap}. Interestingly, in 2D and 3D, all rotations are ``simple rotations", which have a single rotation frequency and rotation plane. Conversely in 4D, the generic type of rotation is a ``double rotation" which has two rotation frequencies and (at least) two orthogonal rotation planes. This leads to vortex structures with no direct analogue in lower dimensions as we have previously begun to explore in Refs.~\cite{mccanna2021} and~\cite{mccannaunequal}, by studying a 4D generalisation of the Gross-Pitaevskii equation (GPE)~\cite{pitaevskii2003}. In particular, we found that equal-frequency double rotation can stabilise a vortex core consisting of two rigid orthogonal planes intersecting at a point~\cite{mccanna2021}, while unequal-frequency double rotation can lead to the formation of unusual skewed and curved vortex surfaces~\cite{mccannaunequal}.

In this paper, we shall combine these research directions to ask if skewed and curved vortex surfaces can also be favoured by equal-frequency double rotation. Our hypothesis is that a 4D superfluid may be able to gain hydrodynamic vortex-vortex interaction energy by having tilted (anti-aligning) vortex planes, albeit at the cost of increased rotational energy. To test this hypothesis, we shall develop a simplified analytical theory building on Ref.~\cite{mccannaunequal}. Numerically, we find that such skewed (and curved) structures can be comparable to or even slightly lower in energy than rigid orthogonal vortex planes~\cite{mccanna2021} for our system parameters.  We also note qualitative similarities between the curvature of these vortex surfaces in 4D and the reconnection of vortex lines in 3D, raising interesting questions for further research. In the future, it will be very interesting to extend our analysis to more experimentally-realistic models for probing higher-dimensional physics~\cite{Kraus_2013,Price_2015,Ozawa2016,Lohse_2018,Zilberberg_2018,sugawa2018second,lu2018topological, kolodrubetz2016measuring, wang2020exceptional,wang2020circuit,Price2018, yu2019genuine,li2019emergence,ezawa2019electric,Weisbrich}, e.g. using approaches such as ``synthetic dimensions"~\cite{Boada2012,Celi2014,Mancini2015,Stuhl2015,Gadway2015, An_2017,Price2017,salerno2019quantized, viebahn2019matter,barbiero2019bose, chalopin2020exploring,Ozawa2016,Yuan2016, ozawa2017synthetic, lustig2019photonic, Yuan2018rings,Yuan2018photonics,Yuan2019, yuan2020creating, Dutt2020,baum2018setting, price2019synthetic,crowley2019half,boyers2020exploring,ozawa2019topological,oliver2021bloch} using which an atomic 4D quantum Hall system has recently been realized experimentally~\cite{bouhiron2022realization}. 

We begin in Section~\ref{sec:4DRot} by reviewing the different types of rotations that occur in different numbers of spatial dimensions, before, in Section~\ref{sec:vortexreview}, briefly summarizing the basic physics of quantum vortices in 2D, 3D and 4D superfluids, as described by the GPE equation. In Section~\ref{sec:Reconnection}, we shall then discuss vortex-vortex reconnection physics in both 3D and 4D. In Section~\ref{sec:Equal}, we consider equal-frequency double rotations, and ask whether this scenario can favour anti-aligning skewed vortex surfaces either analytically and/or numerically.  Finally, in Section~\ref{sec:concl}, we summarize our results and briefly discuss possible future extensions to this work.

\section{Rotations in Different Dimensions}\label{sec:4DRot}

Before discussing the physics of quantum vortices in superfluids, we shall begin by briefly reviewing the different types of rotation that occur in 2D, 3D and 4D systems. For further mathematical details, we refer readers to Ref.~\cite{mccannaunequal} and references there-in. 

In 2D and 3D systems, all rotations are ``simple" meaning that the rotation is fully described by a single rotation angle and the corresponding rotation plane. For example, when represented as a matrix, a rotation of 2D space can be expressed as
\begin{align}
    \begin{pmatrix}
        \cos\alpha & -\sin\alpha \\
        \sin\alpha & \cos\alpha
    \end{pmatrix}, \label{eq:2drot}
\end{align}
where \(\alpha\in (-\pi,\pi]\) is the angle of rotation and where we have defined the origin as the fixed point of the rotation about which all other points are angularly displaced on the rotation plane. Similarly, in 3D, rotations can be expressed in the form
\begin{align}
    \begin{pmatrix}
        \cos\alpha & -\sin\alpha & 0 \\
        \sin\alpha & \cos\alpha & 0 \\
        0 & 0 & 1
    \end{pmatrix},
\end{align}
via a suitable choice of basis, with $\alpha$ again being the rotation angle, and the $x-y$ plane chosen as the rotation plane. The $z$-axis is then the axis of rotation, meaning that it is both the line of points which remain invariant under the rotation and the centre about which the system rotates. From the above form, it is clear that all 3D rotations can be viewed as simple extensions of 2D rotations, in which a third direction is left unchanged. 

Just as 2D rotations can be extended into 3D rotations, so  can the above simple rotations be extended into 4D. Expressed as a matrix this type of rotation can be written (for a suitable basis choice) as
\begin{align}
    \begin{pmatrix}
        \cos\alpha & -\sin\alpha & 0 & 0 \\
        \sin\alpha & \cos\alpha & 0 & 0 \\
        0 & 0 & 1 & 0 \\
        0 & 0 & 0 & 1
    \end{pmatrix}.
    \label{eq:4DSimple}
\end{align}
which again has a single rotation angle $\alpha$ and a single rotation plane (i.e. the $x-y$ plane). However, now instead of the rotation being centered around a point as in 2D or an axis as in 3D, it is centered around a fixed plane (here the $z-w$ plane), which is entirely orthogonal to the rotation plane. 

In contrast to 2D and 3D, the generic type of rotation in 4D are so-called ``double rotations" which are characterised by a single fixed point, and two orthogonal rotation planes, each with their own rotation angle. When represented as a matrix, a suitable basis choice can bring any double rotation into the form
\begin{equation}
    M(\alpha,\beta) =
    \begin{pmatrix}
        \cos\alpha & -\sin\alpha & 0 & 0 \\
        \sin\alpha & \cos\alpha & 0 & 0 \\
        0 & 0 & \cos\beta & -\sin\beta \\
        0 & 0 & \sin\beta & \cos\beta
    \end{pmatrix}, 
    \label{eq:DoubleRot}
\end{equation}
which corresponds to having a rotation angle \(\alpha \in (-\pi,\pi]\) in the \(x\en y\) plane, and a rotation angle \(\beta \in (-\pi,\pi]\) in the \((z\en w)\) plane. As a result, any point on the \(x\en y\) or \(z\en w\) plane will remain on that plane but be rotated around the origin by an angle \(\alpha\) or \(\beta\), respectively. Simple rotations can also be recovered as a special case of double rotations, if either \(\alpha\) or \(\beta\) vanishes.

Another very important special class of double rotations are the so-called ``isoclinic" rotations, in which both rotation angles are equal up to a sign, e.g. \(M(\alpha,\beta)\) with $\beta = \pm \alpha$. These rotations can be classified as being either right-handed or left-handed, depending on the relative senses of rotation in the two planes; for example, \(M(\alpha,\alpha)\) is a left isoclinic rotation of the \(x\en y\) and \(z\en w\) planes, while \(M(\alpha,-\alpha)\) is a right isoclinic rotation of the same planes. It is also known that all left isoclinic rotations commute with all right isoclinic ones, with any rotation of 4D space being decomposable as a product of a left and a right isoclinic rotation \cite{lounesto2001}. 
Another unusual feature of isoclinic rotations is that they have an infinite number of rotation planes, such that any completely orthogonal pair of them can be used as a basis to define a particular isoclinic rotation. This is in contrast to a generic double rotation (i.e. $M (\alpha, \beta)$, with $\alpha \neq \beta \neq 0, \pi$), which have only two unique rotation planes. Interestingly, it is possible to transform between the different rotation planes of a given left (resp. right) isoclinic rotation by applying a suitable right (resp. left) isoclinic rotation to the 4D system~\cite{lounesto2001, mccannaunequal}.   

As we will explain later in Section~\ref{sec:Equal}, we will be particularly interested in coordinate systems related by generic left isoclinic rotations. 
To do so, we will work in double polar coordinates $(r_1,\ta_1,r_2,\ta_2)$ defined by
\begin{equation}
    (x,y,z,w) = (r_1\cos\ta_1,r_1\sin\ta_1,r_2\cos\ta_2,r_2\sin\ta_2). \nonumber
\end{equation}
To transform between the different coordinate systems, it is easiest to work in a complex representation where the position vector is given by \((x+iy,z-iw)^T = (\cpolar{1},\ccpolar{2})^T\). In this case, the general left isoclinic rotation going from unprimed to primed coordinate systems can then be shown to be~\cite{mccannaunequal}
\begin{align}
    \begin{pmatrix}
        \cpolar[']{1} \\
        \ccpolar[']{2}
    \end{pmatrix}
    &= 
    \begin{pmatrix}
        \cos\eta & e^{i\varphi}\sin\eta \\
        -e^{-i\varphi}\sin\eta & \cos\eta
    \end{pmatrix}
    \begin{pmatrix}
        \cpolar{1} \\
        \ccpolar{2}
    \end{pmatrix},
    \label{eq:RotPlaneTransformation}
\end{align}
which depends only on two parameters, \(\eta\in[0,\pi/2]\) and \(\varphi\in[0,2\pi)\), and where all redundant parameters have already been removed. Physically, the parameter $\eta$ denotes the tilt angle between the planes $r_{1}=0$ and $r'_{1}=0$, while $\varphi$ represents the direction of this tilt.

\section{Review of superfluid vortices} \label{sec:vortexreview}

Having summarized the main features of rotations in different numbers of spatial dimensions, we shall now briefly review how these rotations can stabilise different types of quantized superfluid vortices in 2D, 3D and 4D systems. Throughout, we shall be considering a system of weakly-interacting bosons as described by the time-independent Gross-Pitaevskii equation~\cite{pitaevskii2003}
\begin{equation}
	-\frac{\hbar^2}{2m}\nabla^2\psi + g|\psi|^2\psi = \mu\psi,
	\label{eq:GPE}
\end{equation}
where \(m\) is the particle mass, $g$ is the strength of interactions, \(\mu\) is the chemical potential and where we have assumed there is no external trapping potential. Here $\psi$ is the complex order parameter, which is a function of all spatial dimensions in the system, and from which we can define the superfluid density as $\rho = |\psi|^2$, the superfluid phase as $S = \Arg\psi$, and the superfluid velocity as $\mathbf{v} = \frac{\hbar}{m}\nabla S$~\cite{pitaevskii2003}. Crucially, it is straightforward to show that this form of the velocity field implies that the superfluid circulation must be quantized, and hence that a simply-connected superfluid cannot rotate as a rigid body, but instead forms quantized vortices~\cite{pitaevskii2003, cooper2008rapidly, fetter2009, pethick2002, madison2000, madison2001, matthews1999,abo-shaeer2001}.

\subsection{Superfluid Vortices in 2D and 3D}
\label{sec:2DVortex}

In 2D, a superfluid vortex consists of a point-like vortex core where the density goes to zero, and around which the superfluid phase winds by a quantized amounts~\cite{pitaevskii2003, cooper2008rapidly, fetter2009, pethick2002}. Mathematically, a rotationally-symmetric 2D vortex in the $x-y$ plane can be described by 
\begin{equation}
    \psi (r,\ta) = |\psi(r)| e^{i k \ta}, 
    \label{eq:2D}
\end{equation}
where the vortex core lies at the origin, $r=0$. Using dimensionless units, we can denote the density profile as $|\psi(r)|= f_k(r)$, which is a real-valued function that vanishes towards the vortex core and which can be obtained by solving the (dimensionless) 2D GPE equation numerically. The topological integer $k$ in Eq.~\eqref{eq:2D} corresponds to the quantized phase winding of the superfluid around the vortex core, meaning that the superfluid velocity is then given by~\cite{pitaevskii2003}
\begin{equation}
{\bf v} = \frac{\hbar}{m} \nabla  (k  \ta) =  \frac{\hbar}{m r} k  \bfhat{\ta},   
\label{eq:velocity}
\end{equation}
where $\bfhat{\ta}$ is the unit vector pointing along the $\ta$ direction. 

Energetically, vortices can be introduced by rotating the system (or equivalently by engineering an artificial magnetic field)~\cite{fetter2009,cooper2008rapidly}. In a rotating reference frame, the GPE [Eq.~\eqref{eq:GPE}] becomes
\begin{equation}
	\left[-\frac{\hbar^2}{2m}\nabla^2 + g|\psi|^2 - \mathbf{\Omega}\cdot\hat{\mathbf{L}}\right] \psi = \mu\psi, \label{eq:GPE2DR}
\end{equation}
 where $\mathbf{\Omega}$ is the vector of rotation frequencies and $\hat{\mathbf{L}} = -i\hbar\mathbf{r}\times\nabla$ is the (3D) angular momentum operator~\cite{pitaevskii2003}. For a rotating 2D system, $\mathbf{\Omega}= \Omega \bfhat{z}$, corresponding to 2D simple rotation, i.e. Eq~\eqref{eq:2drot} with $\alpha = \Omega t$. As a result, the energy of the system is reduced by $\Delta E_{\text{rot}} = \Omega \langle \hat{L}_z \rangle$, where $\langle \hat{L}_z \rangle$ is the expectation value of the angular momentum operator, $\hat{L}_z$, with respect to the order parameter. As vortices carry a finite amount of angular momentum, rotation can therefore energetically stabilise a vortex, provided that this energy reduction outweighs the energy costs~\cite{pethick2002, pitaevskii2003}. This occurs above a critical frequency, $\Omega_c^{2D}$, which, for a superfluid in a disc of radius $R$ with hard-wall boundary conditions (and no additional external potentials), can be estimated as
\begin{equation}
\Omega_c^{2D} = k \frac{\hbar }{m R^2}   \ln \left( 2.07 \frac{R}{\xi}\right), 
\label{eq:2Dcrit}
\end{equation}
where $\xi$ is the healing length, which satisfies $\hbar^2/m\xi^2 = gn = \mu$, with $n$ being the uniform background superfluid density.  (Note that often a factor of $1/2$ is included in the definition of $\xi$~\cite{pitaevskii2003}.) This shows that as the rotation frequency is increased from zero, the ground state will first change from a state with no vortices to a state with a single vortex with $k=1$~\cite{pethick2002}. 

At even higher frequencies, a straightforward analysis of the various energy contributions predicts that it is always energetically unfavourable (without additional potentials) to create a multiply charged vortex with $|k| >1$, rather than multiple singly-charged vortices with $|k|=1$~\cite{pitaevskii2003}. As the rotation frequency increases, the ground state therefore changes from a state with a single $k=1$ vortex to a state with two $k=1$ vortices and so on. Considering multiple vortices in a system, it can then be shown that vortices with winding numbers of the same sign will interact repulsively, while those with opposite sign (i.e. a vortex and anti-vortex pair) will interact attractively~\cite{pethick2002}.

We can straightforwardly generalise the above description to vortices in a 3D superfluid~\cite{pitaevskii2003, pethick2002, verhelst2017vortex}. In this case, a superfluid vortex consists of a vortex core, which can be approximated as an extended 1D line. This core must either begin and end on the surface of the system, in which case they are referred to as ``vortex lines" or ``vortex filaments", or else form a closed loop within the 3D superfluid, in which case they are called ``vortex rings"~\cite{verhelst2017vortex,Anderson2001,rosenbusch2002dynamics,komineas2007vortex,carretero2008nonlinear, bisset2015robust, wang2017single}. For the purposes of this paper, we will focus on vortex lines as we are studying the lowest-energy structures that can be stabilised by rotation; however, it would be very interesting in the future to also consider how vortex rings can be generalised to higher dimensions. Mathematically, a cylindrically-symmetric vortex line in 3D can be described~\cite{verhelst2017vortex} e.g. by
\begin{equation}
        \psi (r, \ta, z) = |\psi(r, z)| e^{i k \ta}, 
\end{equation}
in cylindrical polar coordinates $(r, \ta, z)$, where the rotation axis lies in the $z$ direction. Without additional potentials, the density profile in dimensionless units is then given by $|\psi(r, z)| = f_k (r)$, which is independent of $z$ and where $ f_k (r)$ is the real-valued function found by numerically solving the 2D (dimensionless) GPE. As a result, in this simplest case, a 3D vortex line has the same velocity field and the same critical frequency (in a cylindrical system) as a 2D vortex~\cite{verhelst2017vortex}.

\subsection{Superfluid Vortices in 4D}
\label{sec:4Dsuper}

We will now briefly review the different types of vortex structures that we previously identified as low-energy states under rotation [Eq.~\eqref{eq:GPE}] in Refs.~\cite{mccanna2021} and~\cite{mccannaunequal}. Throughout, we will focus on the 4D GPE as it is both the most natural and simple model in which to study how superfluid vortex structures would be affected by extra dimensions, and it is also plausible as a mathematical description of low-temperature interacting bosons moving in a hypothetical universe of four spatial dimensions~\cite{Wodkiewicz,Stampfer,Trang,mccanna2021}. However, as discussed in Ref.~\cite{mccannaunequal} and briefly in our conclusions in Section~\ref{sec:concl}, it will be interesting and important in the future to generalise these results to a more realistic experimental model that is based e.g. on adding one or more ``synthetic dimensions" to ultracold atoms~\cite{Boada2012,Celi2014,Mancini2015,Stuhl2015,Gadway2015, Price_2015,An_2017,Price2017,salerno2019quantized, viebahn2019matter,barbiero2019bose, chalopin2020exploring,oliver2021bloch, bouhiron2022realization}. 

Before proceeding, it is important to note that the rotating frame GPE [Eq~\eqref{eq:GPE2DR}] depends on the angular momentum operator, which in 2D and 3D can be represented as a vector. However, in 4D there are six Cartesian coordinate planes, meaning that the rotation group \(SO(4)\) of four-dimensional space has {\it six} generators that physically describe angular momentum. Hence, the angular momentum operator in 4D  is a 4x4 antisymmetric tensor, with components $\hat{L
}_{\gamma \delta}$, which correspond to the angular momentum in the $\gamma\en \delta$ plane, with ${\gamma, \delta} \in \{x, y, z, w\}$. Using this notation, the 4D GPE then takes the form
\begin{equation}
	\left[-\frac{\hbar^2}{2m}\nabla^2 + g|\psi|^2 - \sum_{ \gamma \delta} \Omega_{\gamma \delta}\hat{L
}_{\gamma \delta}  \right] \psi = \mu\psi, \label{eq:GPE4DR}
\end{equation}
where $\Omega_{\gamma \delta}$ is the rotation frequency for the $\gamma\en \delta$ plane. In the following, we will focus on two possibilities: the first is that of simple rotation in 4D, which is characterised by a single rotation frequency, e.g. $\Omega_{xy} \equiv \Omega \neq 0$. This can be understood as a usual 2D or 3D rotation extended into a fourth dimension, and corresponds to taking $\alpha = \Omega t $ in Eq~\eqref{eq:4DSimple}. The second, and our main focus in this work, is that of 4D double rotation, where the rotating-frame GPE can be written as
\begin{equation}
	\left[-\frac{\hbar^2}{2m}\nabla^2 + g|\psi|^2 - \Omega_1 \hat{L}_1 - \Omega_2 \hat{L}_2\right] \psi = \mu\psi,
	\label{eq:GPER}
\end{equation}
where $\Omega_j$ is the rotation frequency and $\hat{L}_j$ is the angular momentum operator in plane $j=1,2$. In this paper, we shall choose plane 1 to be the $x\en y$ plane (i.e. $\Omega_1 \equiv \Omega_{x y}$, $\hat{L}_1 \equiv \hat{L}_{xy} $), and plane 2 to be the $z\en w$ plane (i.e. $\Omega_2 \equiv \Omega_{z w}$, $\hat{L}_2 \equiv \hat{L}_{zw}$). This corresponds to taking $\alpha = \Omega_1 t $ and  $\beta = \Omega_2 t $ in Eq~\eqref{eq:DoubleRot}. (Note that this scenario is closely related to certain types of 4D quantum Hall models in which magnetic fields in two completely orthogonal planes are used to generate a nontrivial second Chern number \cite{Price_2015,Ozawa2016,Lohse_2018,Zilberberg_2018,Mochol_Grzelak_2018}.)

\subsubsection{A Single Vortex Plane}

Just as a 3D vortex line is an extension of a 2D vortex point, so can we consider a rigid 4D ``vortex plane", in which the vortex core has become extended to cover an entire 2D plane within the superfluid~\cite{mccanna2021}. In dimensionless units, the corresponding order parameter can be described e.g. by:
\begin{equation}
    \psi (r_1, \ta_1, r_2, \ta_2) =  f_k(r_1)e^{ik\ta_1},
    \label{eq:4Dsingle}
\end{equation}
where $(r_1,\ta_1, r_2, \ta_2)$ are the double polar coordinates introduced previously. Without additional potentials, the density profile, $f_k(r)$, is again the radial function found numerically from the 2D GPE, and so the vortex core spans the entire $z\en w$ plane. Physically, this is the most natural extension of vortices from 2D and 3D into 4D, as the extra dimension, $w$, plays no role in either the rotation or in the form of the order parameter. It can therefore also be expected that such a vortex plane can be energetically stabilised by a 4D simple rotation, e.g. as described above with $\Omega_{xy} \neq 0$ and all other rotation frequencies being equal to zero, as has been verified numerically in Ref.~\cite{mccanna2021}.

We can also consider a single rigid vortex plane under 4D double rotation [c.f. Eq.~\eqref{eq:GPER}]. First of all, in the limit that one of the two rotation frequencies is very small, the scenario remains close to that of simple rotation, and we expect that a single vortex plane will  be energetically favoured (provided the other rotation frequency is sufficiently big)~\cite{mccanna2021}. More generally, it can be shown that the vortex plane will tilt so as to fully align with whichever plane has the higher rotation frequency, so as to fully benefit from the energy reduction due to rotation~\cite{mccannaunequal}. However, if the two rotation frequencies both become  large enough, then a single vortex plane is unlikely to be the ground-state over a significant frequency range, as other more favoured structures, based on pairs of vortex planes, can emerge, as we shall now review~\cite{mccanna2021, mccannaunequal}.

\subsubsection{A Pair of Orthogonal Vortex Planes}

Under double rotation, the 4D GPE [Eq~\eqref{eq:GPER}] depends on two commuting operators $\hat{L}_1$ and $\hat{L}_2$, suggesting that we look for simultaneous eigenstates of both angular momentum operators. Physically, this suggests the possibility of having vortex structures composed of a pair of completely orthogonal vortex planes that intersect at the origin. Such a structure can be described, e.g. by the following ansatz for the dimensionless order parameter~\cite{mccanna2021}
\begin{equation}
    \psi (r_1, \ta_1, r_2, \ta_2) = f_{k_1, k_2} (r_1, r_2)e^{i k_1 \ta_1 + ik_2 \ta_2},
    \label{eq:OrthogonalAnsatz}
\end{equation}
where $k_1$ and $k_2$ are the integer winding numbers in the two rotation planes, and the structure is hence characterised by \(\mathbb{Z}\times \mathbb{Z}\) topological invariants. The real-valued function, $f_{k_1, k_2} (r_1, r_1)$, denotes the 4D superfluid density profile, which we assume only depends on the radial coordinate in each rotation plane. The form of this function can be found by solving the (dimensionless) GPE numerically; doing so shows that the resulting function is close to a product of 2D vortex profiles, i.e. $f_{k_1, k_2} (r_1, r_1) \approx f_{k_1} (r_1) f_{k_2} (r_2)$, although this approximation breaks down, due to the intrinsic nonlinearity of the GPE, near the origin where the vortex planes intersect~\cite{mccanna2021}. From Eq.~\eqref{eq:OrthogonalAnsatz}, it can also be straightforwardly seen that the associated superfluid velocity field is given by~\cite{mccanna2021}
\begin{equation}
{\bf v} = {\bf v}_1 + {\bf v}_2 =  \frac{\hbar}{m} \left( \frac{k_1}{r_1}  \bfhat{\ta}_1 + \frac{k_2}{r_2}  \bfhat{\ta}_2 \right),   
\label{eq:velocity4D}
\end{equation}
corresponding to a superposition of the velocity field of a 2D vortex in each rotation plane [c.f. Eq.~\eqref{eq:velocity}]. 

Interestingly, a pair of orthogonal vortex planes is favoured over a single rigid plane [Eq.~\eqref{eq:4Dsingle}] by sufficiently strong equal-frequency double rotation, i.e. $\Omega_1 = \Omega_2 = \Omega$, as we showed numerically in Ref.~\cite{mccanna2021,mccannaunequal}. This is because, although a pair of planes has a higher intrinsic (e.g. hydrodynamic) energy cost than a single plane, such a structure can benefit from a much greater energy reduction under equal-frequency rotation due to having angular momentum in both rotation planes simultaneously [c.f. Eq~\eqref{eq:GPER}]. Similar to in 2D and 3D, it is also possible to estimate a critical frequency 
above which the simplest pair of orthogonal vortex planes (e.g. with $|k_{1,2}|=1$) becomes lower energy than the uniform state with no vortices. For a superfluid in a 4D  hypersphere (or ``4D ball") geometry, which has hard-wall boundaries at $r_1^2 + r_2^2 = R^2$ with $R$ being the hyperspherical radius, this critical velocity is approximately given by~\cite{mccannaunequal}
\begin{align}
    \Omega_c \approx 2\frac{\hbar}{mR^2}\ln\left(2.07\frac{R}{\xi}\right).
    \label{eq:4Dcrit}
\end{align}
This shall be used in the following to report frequencies in units of \(\Omega_c\). (Note that Ref.~\cite{mccanna2021} worked in units of \(\Omega_c^{2D}\) [Eq~\eqref{eq:2Dcrit}], but these are simply related as \(\Omega_c = 2\Omega_c^{2D}\).) It should also be noted that if the rotation frequencies become high enough compared to $\Omega_c$, then we expect that it will become energetically favourable to introduce many vortices or more complicated vortex structures; however, this goes beyond the scope of our work. 

Finally, while the above ansatz for orthogonal vortex plane [Eq.~\eqref{eq:OrthogonalAnsatz}] picks out the \(x\en y\) and \(z\en w\) planes preferentially, equal-frequency double rotations are examples of isoclinic rotations [c.f. Section~\ref{sec:4DRot}], and hence have an infinite number of rotation planes~\cite{lounesto2001, mccannaunequal}. This means that a different but equally suitable ansatz could have been defined with respect to any orthogonal pair of these planes, provided that the boundary conditions respect this symmetry, e.g. as is the case for the 4D ball geometry. In practice, numerical calculations are carried out on a discretized Cartesian grid and for an initial state that breaks the isoclinic symmetry [see Appendix~\ref{app:methods}], meaning that this degeneracy is not typically reflected in numerical results. 

\subsubsection{A Pair of Non-orthogonal Vortex Planes}

In this paper, we want to explore whether equal-frequency double rotations can stabilise a different type of vortex structure that is composed of a pair of {\it non-orthogonal} vortex planes. This is inspired by work in Ref.~\cite{mccannaunequal}, where we showed that a suitably skewed pair of rigid vortex planes could have lower energy than an orthogonal pair of planes, under unequal-frequency double rotation. In preparation, we shall therefore now briefly introduce this type of vortex structure.  

Following Ref.~\cite{mccannaunequal}, an example ansatz for the dimensionless order parameter of a pair of non-orthogonal, rigid vortex planes is given by
\begin{align}
    &\psi = r_1^{|k_1|}{r_2'}^{|k_2|}e^{i\left(k_1\ta_1+k_2\ta'_2\right)}g(r_1^2,r_2'^2),
    \nonumber \\
    &= \left(x+\sigma_1iy\right)^{|k_1|} \left(z'+ \sigma_2iw'\right)^{|k_2|} g(x^2+y^2,z'^2+w'^2), \qquad
    \label{eq:skewVortices}
\end{align}
where the primed coordinate system is tilted with respect to the unprimed coordinate system, as related by a general double rotation. Physically, this ansatz describes a pair of vortex planes along \(x=y=0\) and \(z'=w'=0\) respectively, which intersect at the origin. Here, \(k_{1,2}\) are the winding numbers of the two vortices and \(\sigma_j=\sgn(k_j)\). The function \(g\) is given by \(g(r_1^2,r_2'^2)=\text{const} \times \ f_{k_1,k_2}(r_1,r_2')/r_1^{|k_1|}{r_2'}^{|k_2|}\), where \(f_{k1,k2}\) is the dimensionless profile from Eq.~\eqref{eq:OrthogonalAnsatz}~\cite{mccannaunequal}. 

Through an appropriate choice of basis, the relationship between the primed and unprimed coordinates can be taken without loss of generality as~\cite{mccannaunequal} 
\begin{equation}
    z' = \sin\alpha_1x + \cos\alpha_1z, \qquad
    w' = \sin\alpha_2y + \cos\alpha_2w.
    \label{eq:primedZW}
\end{equation}
with \(\alpha_{1,2} \in [0,\pi/2)\). (Note that a pair of orthogonal vortex planes, as discussed in the previous section, would correspond to taking \(\alpha_1=\alpha_2=0\).) Given again a spherically-symmetric 4D superfluid of radius \(R\), such that \(r_1^2+r_2^2=r_1'^2+r_2'^2\leq R\), we will assume that the velocity fields induced by each vortex have the following simple forms
\begin{equation}
    \mathbf{v}_1 = k_1 \vDimens \frac{\bfhat{\ta}_1}{r_1} \qquad \mathbf{v}_2' = k_2 \vDimens \frac{\bfhat{\ta}_2'}{r_2'}.
    \label{eq:4DVelocity}
\end{equation}
Note that, in general, these velocity fields are not orthogonal to each other, and therefore there can be a non-zero hydrodynamic vortex-vortex interaction between the two planes~\cite{mccannaunequal}. We shall discuss the form of this hydrodynamic interaction in more detail in Sec.~\ref{sec:Equal}. However, to make a simple argument, we can recall that, as mentioned above, 2D vortices with winding numbers of the same sign, and hence velocity fields circulating in the same sense, will interact repulsively. Then, intuitively in 4D, if the second vortex plane is tilted such that its velocity field begins to align (resp. anti-align) with that of the first vortex plane, we can expect there to be an energetic cost (resp. benefit) due to the effectively repulsive (resp. attractive) interaction between the pair of planes~\cite{mccannaunequal}.   

As an example, we can consider the special case in which the double rotation between the primed and unprimed coordinates is isoclinic. This means that \(\alpha_2 = \nu\alpha_1\), with \(\nu = \pm1\) denoting if the rotation is left (\(-\)) or right (\(+\)) isoclinic respectively. If we then define \(\eta\equiv\alpha_1\) for simplicity, we see that the primed coordinates [Eq.~\eqref{eq:primedZW}] can now be expressed as
\begin{equation}
    z'+iw' = c(z+iw) + s(x+\nu iy),
    \label{eq:isoTilt}
\end{equation}
where we have applied the shorthand \(c=\cos\eta\), \(s=\sin\eta\). In this case, the ansatz for non-orthogonal intersecting vortex planes becomes~\cite{mccannaunequal}
\begin{align}
    \psi = (x+\sigma_1iy)^{|k_1|}\left[c(z+\sigma_2iw) + s(x+\nu\sigma_2 iy)\right]^{|k_2|}g,
\end{align}
where the arguments of \(g\) have been suppressed for brevity. Then if \(\nu=\sigma_1\sigma_2\), we have \(x+\sigma_1iy = x+\nu\sigma_2iy\), and the planes are skewed so as to begin to perfectly align, while if \(\nu=-\sigma_1\sigma_2\), they anti-align. This in turn suggests that \(\nu=\sgn(k_1k_2)\) gives rise to a repulsive interaction between the planes, while \(\nu=-\sgn(k_1k_2)\) will lead to an attractive interaction~\cite{mccannaunequal}. 

In Ref.~\cite{mccannaunequal}, we found that such a pair of non-orthogonal vortex planes could be lower in energy than the corresponding orthogonal configuration for unequal-frequency double rotation. This result can be understood intuitively by 
noting that, for non-orthogonal planes, there is a competition between the rotational energy [c.f. Eq.~\eqref{eq:GPE4DR}] and the interaction energy. In particular, as one of the rotation frequencies is larger than the other (e.g. $\Omega_1 > \Omega_2$), the system can benefit energetically by tilting the vortex planes so that they have increased angular momentum in the rotation plane with the higher frequency; however, this means that the vortex planes have also tilted towards each other in an aligning sense, and hence will interact repulsively. Balancing these two energetic considerations led us to predict optimal tilt angles for the two planes, which was found to be in good agreement with numerics~\cite{mccannaunequal}. In this paper, we will carry out analogous analytical and numerical calculations for the case of equal-frequency double rotation.

\section{Vortex Reconnection}
\label{sec:Reconnection}

Before proceeding, it is interesting to note that the tilted non-orthogonal vortex planes found numerically in Ref.~\cite{mccannaunequal} exhibited an avoided crossing near the origin, i.e. instead of an intersection at the origin as in the ansatz [Eq.~\eqref{eq:skewVortices}] or as found numerically for orthogonal vortex planes~\cite{mccanna2021}. As we shall show in Sec.~\ref{sec:Equal}, we numerically observe a similar avoided-crossing phenomenon also for equal-frequency double rotation. Moreover, the shape of these curved surfaces is reminiscent of the shape of 3D vortex lines shortly after a reconnection event; this motivates us to first review briefly how reconnections can be approximated analytically in 3D superfluids~\cite{nazarenko2003}, before presenting how this theory can be extended to 4D. This will serve as a comparison for numerical results in Sec.~\ref{sec:Equal}. 

\subsection{Reconnections in 3D}

In 3D, it is well-known that when two vortex lines are made to intersect, they will dynamically reconnect and move apart so as to remove the intersection point~\cite{schwarz1988three}. This process plays an important role, e.g. in quantum turbulence~\cite{vinen2002quantum}, and has been studied in detail theoretically~\cite{koplik1993vortex,nazarenko2003,tebbs2011approach, zuccher2012quantum,Scheeler2014, allen2014vortex, villois2017universal,villois2020}, and in various superfluid experiments~\cite{bewley2008characterization, Anderson2001, serafini_2015,serafini2017}. 

At very short times before or after the moment of reconnection, Nazarenko and West showed in a seminal paper that it is possible to analytically approximate the wavefunction solution close to the reconnection point, by assuming that nonlinear effects are small~\cite{nazarenko2003}. This assumption is justified by noting that the wave-function is spatially continuous and vanishes at the vortex cores, meaning that the particle density is low near the reconnection point. To a first approximation, the reconnection evolution can then be described by the (dimensionless) 3D linearised time-dependent GPE, i.e. the Schr{\"o}dinger equation
\begin{equation}
- \frac{1}{2} \nabla^2 \psi  = i \dot{\psi},
    \label{eq:Schrodinger}
\end{equation}
close to the intersection point (at ${\bf r}=0$). If the wave-function at the moment of intersection ($t=t_0$) is denoted by $\psi = \psi_0$, then after a short time interval $\Delta t = (t-t_0)$, the state evolution can be approximated as:
\begin{equation}
\psi = \psi_0 + i \frac{\Delta t }{2} \nabla^2 \psi_0 .  \label{eq:timeev}
\end{equation}
At $t=t_0$, a reasonable ansatz for the wave-function near the intersection point is given by~\cite{nazarenko2003}
\begin{eqnarray}
\psi_0  = z  + i (az + b x^2 - cy^2),
\end{eqnarray}
where $a, b, c$ are some positive constants, corresponding to a state containing two straight vortex cores (defined by $\psi_0 =0$) that intersect at the origin. According to Eq.~\eqref{eq:timeev}, such a state will evolve dynamically to~\cite{nazarenko2003}
\begin{equation}
\psi = z - (b-c) \Delta t + i (az + b x^2 - cy^2) . 
\end{equation}
For times both before ($\Delta t <0$) or after ($\Delta t >0$) the moment of reconnection, this solution describes two unconnected hyperbolae, corresponding physically to two separated and curving vortex filaments. Interestingly, even such a simple linear approach reproduces many of the observed properties of vortex reconnections found in numerical simulations, such as that the vortex lines locally approach each other in an anti-parallel configuration~\cite{nazarenko2003, villois2017universal}. However, as this method does not describe the behaviour far from the vortex cores, the linear approach cannot predict effects such as the far-field emission of sound waves by the reconnection event; to overcome such limitations, the linear solution can be substituted back into the GPE including nonlinear terms in order to find successive nonlinear corrections, and hence to analytically calculate a fully nonlinear analytical solution within a finite volume and finite evolution time~\cite{nazarenko2003}.    

\subsection{Reconnections in 4D}

Here we show that non-orthogonal vortex planes in 4D generically do not form a stationary state. In particular, we show that an initial state containing two non-orthogonal singly charged vortex planes that intersect at a point will undergo reconnection. As we shall see, in contrast to reconnections of extended vortex lines in 3D~\cite{nazarenko2003}, the vortex core remains a single connected object at all times during this four-dimensional reconnection. (We note that a single connected core can arise in evolved states of the 3D GPE when considering self-intersection and reconnection of vortex loops~\cite{andryushchenko2018} or of a pair of linked vortex loops~\cite{villois2017universal,villois2020,Proment2020}.) 

To proceed, we follow the same analysis as that described above; close to a vortex core  the density is small, so for short times we may describe the evolution of a vortex core with the linearised dimensionless GPE [Eq~\eqref{eq:Schrodinger}] now in 4D. 
We will take our ansatz for non-orthogonal vortex planes [Eq~\eqref{eq:skewVortices}] as an initial state, with \(|k_1|=|k_2|=1\), assuming an idealised case of an infinite condensate that is homogeneous away from the vortex core. Looking at the immediate vicinity of the intersection point between the planes, such that \(r_1\) and \(r_2'\) are both small (compared to \(\xi\)), we can approximate the function \(g\) in Eq~\eqref{eq:skewVortices} to leading order as \(g(0,0)\). Any constants can then be divided out of the linear evolution [Eq~\eqref{eq:Schrodinger}], so in the immediate vicinity of either vortex core we can approximate the state as
\begin{equation}
    \psi_0 = (x+iy)(z'+\sigma iw'),
\end{equation}
where we have assumed \(\sigma_1=1\) without loss of generality, and \(\sigma=\sigma_2\) then denotes the relative sign of \(k_1k_2\), and hence the relative orientation of the two planes.
Substituting the equation for \(z'\) and \(w'\) [Eq~\eqref{eq:primedZW}] into our initial condition gives
\begin{equation}
    \psi_0 = \sin\alpha_1 x^2 - \sigma\sin\alpha_2 y^2 + \text{cross-terms},
    \label{eq:initUnprimed}
\end{equation}
where we have suppressed the cross-terms since they will not contribute in what follows. If we let this initial state undergo the Schrodinger evolution of Eq~\eqref{eq:Schrodinger}, then after a short time \(\Delta t\) the evolved state is given by Eq.~\eqref{eq:timeev} as in 3D. 
The Laplacian of the initial state contains only contributions from the first two terms in Eq~\eqref{eq:initUnprimed}, and is given by
\begin{equation}
    \nabla^2\psi_0 = 2(\sin\alpha_1-\sigma\sin\alpha_2),
\end{equation}
such that the evolved state is given by
\begin{equation}
    \psi = (x+iy)(z'+\sigma iw') + i\Delta t(\sin\alpha_1-\sigma\sin\alpha_2).
    \label{eq:reconnection}
\end{equation}
Note that at the point of reconnection \(\Delta t=0\), the vortex core is given by two non-orthogonal planes intersecting at a point. 
Interestingly this is also true at all times if the two planes are related by an isoclinic rotation, such that \(\alpha_1=\nu\alpha_2\) -- provided that \(\nu = \sigma\). In this case the linearised equation [Eq~\eqref{eq:timeev}] predicts no dynamics and hence no reconnection. This corresponds to a right iscoclinic rotation for \(\nu=\sigma=1\) and a left handed one for \(\nu=\sigma=-1\). In both of these cases, however, we can use the definition of the primed coordinates [Eq~\eqref{eq:isoTilt}] to see that, when \(\nu=\sigma\)
\begin{align}
    z'+\sigma iw' = c(z+\sigma iw) + s(x+iy),
\end{align}
such that the planes are skewed in a purely aligning sense, meaning that their interaction is repulsive~\cite{mccannaunequal}. This raises the possibility that such non-orthogonal purely aligning planes could form a stationary state; however, we expect that this is not true for the full nonlinear dynamics that apply at larger distances from the core. This is because the double rotation relating the primed and unprimed coordinates affects the function \(g\) in Eq.~\eqref{eq:skewVortices} even given the form we have assumed for it. It would be interesting in future work to test this hypothesis by applying the nonlinear analytics used for the reconnection of vortex lines in 3D~\cite{nazarenko2003} to these planes in 4D. Regardless, we can speculate that for purely aligning skewed planes to reconnect in the same way as the general skewed planes, the core would have to first twist near the intersection point so that --- in the immediate vicinity of this point --- the core forms two planes that are skewed with some anti-alignment component (\(\alpha_1\neq\sigma\alpha_2\)). This is reminiscent of the case in 3D, where it it well known that vortex lines are always anti-aligning very close to the reconnection point \cite{nazarenko2003}.

\begin{figure}[!]
    \centering        \includegraphics[width=0.48\textwidth]{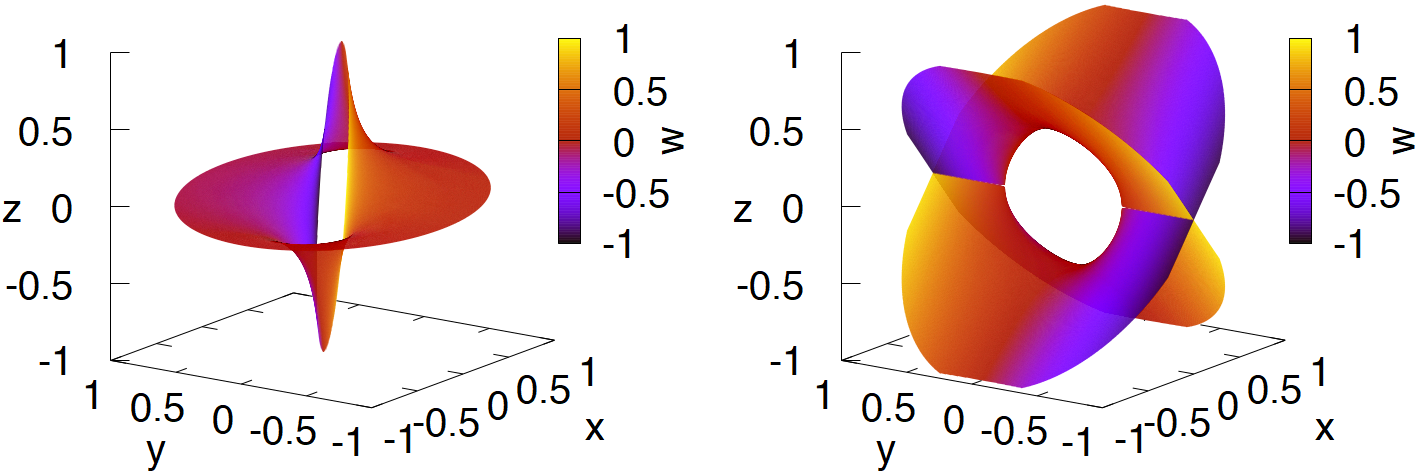}
    \caption{Perturbed orthogonal vortex planes given by the zeroes of Eq~\eqref{eq:OrthoReconnected}, with \(a=0.1,\) \(b=0\). All quantities here are dimensionless. (Left) The vortex core in the original basis, plotted using the height (\(z(x,y)\)) and colour (\(w(x,y)\)) functions given in Eqs~\eqref{eq:OrthoRecCore}. The intersection point has become an avoided crossing, and the vortex core approaches each of the original planes as the distance from the origin increases. (Right) The same state after a rotation of the coordinates given by Eq~\eqref{eq:CrossedRotation}, with \(z(x,y)\) and \(w(x,y)\) now defined by Eqs~\eqref{eq:OrthoRecCoreCrossed}.}
    \label{fig:OrthoReconnected}
\end{figure}

To get an idea of what the reconnected core structure looks like, consider a simpler state given by
\begin{equation}
    \psi_{\perp} = (x+iy)(z+iw) - a - ib
    \label{eq:OrthoReconnected}
\end{equation}
which describes a pair of completely orthogonal planes perturbed by a complex constant, in the same way that Eq~\eqref{eq:reconnection} describes a pair of non-orthogonal planes perturbed by the last term which is a constant for a given value of \(\Delta t\). Note that the equation of the core, \(\psi_\perp = 0\), has the form of the equation of a hyperbola, but with complex numbers instead of real numbers. This is algebraically reminiscent of reconnections in 3D, but there are some crucial differences. For example, this equation now describes a single connected surface rather than two disconnected curves.

The advantage of Eq~\eqref{eq:OrthoReconnected} over Eq~\eqref{eq:reconnection} is that it allows us to focus on the effect of the perturbation without the complication of the skewed planes and the time dependence of the state. The location of any vortex cores is given by the set of zeroes of the order parameter, which in this simpler case are all the points which satisfy
\begin{align}
    xz - yw = a, \qquad
    xw + yz = b.
    \label{eq:reconnectedImplicit}
\end{align}
Solving these equations for \(z = z(x,y)\) and \(w = w(x,y)\) gives us
\begin{align}
    z = \frac{ax + by}{x^2 + y^2}, \qquad
    w = \frac{bx - ay}{x^2 + y^2},
    \label{eq:OrthoRecCore}
\end{align}
which can be plotted as a surface in 3+1 dimensions, where the extra \(w\) dimension is given by colour. This is shown in the left panel of Fig~\ref{fig:OrthoReconnected} for \(a=0.1\), \(b=0\). This figure shows that there is no longer an intersection at the origin, although the structure around this point is difficult to make out. We can get a clearer view by performing a double rotation, and looking at this object from a different perspective. In particular, rotating our coordinates according to
\begin{equation}
    \begin{pmatrix}
        x \\ y \\ z \\ w
    \end{pmatrix} \to
    \frac{1}{\sqrt{2}}
    \begin{pmatrix}
        1 & 0 & 1 & 0 \\
        0 & 1 & 0 & 1 \\
        -1 & 0 & 1 & 0 \\
        0 & -1 & 0 & 1
    \end{pmatrix}
    \begin{pmatrix}
        x \\ y \\ z \\ w
    \end{pmatrix}
    \label{eq:CrossedRotation}
\end{equation}
allows us to visualize both planes at the same time. This transforms 
Eq~\eqref{eq:OrthoReconnected} into
\begin{align}
    \psi_\perp = \frac{1}{2}(z+iw)^2 - \frac{1}{2}(x+iy)^2 -a - ib,
\end{align}
and Eqs~\eqref{eq:reconnectedImplicit} into
\begin{align}
    -x^2 + y^2 + z^2 - w^2 = 2a, \qquad
    -xy + zw = b,
\end{align}
which can again be solved for \(z(x,y)\) and \(w(x,y)\), giving
\begin{align}
    \begin{aligned}
        &z^2 = \left[ A(x,y)^2 + B(x,y)^2 \right]^{\frac{1}{2}} + A(x,y), \\
        &w^2 = \left[ A(x,y)^2 + B(x,y)^2 \right]^{\frac{1}{2}} - A(x,y),
        \label{eq:OrthoRecCoreCrossed}
    \end{aligned}
\end{align}
where \(A(x,y)= a + (x^2-y^2)/2\), and \(B(x,y) = b + xy\). These (now two-branched) solutions are plotted, for \(a=0.1\), \(b=0\) in the right panel of Fig~\ref{fig:OrthoReconnected}, giving a clearer view of the core structure near the origin. Again, we see that the intersection point has been replaced by a kind of avoided crossing reminiscent of the reconnection of intersecting vortex lines in 3D, but where the vortex core remains a single connected 2D region. Recall that the perturbed orthogonal state [Eq~\eqref{eq:OrthoReconnected}] is a simplification of the perturbed skewed state undergoing linear dynamical reconnection [Eq~\eqref{eq:OrthoRecCore}]. To visualise this non-orthogonal reconnecting state for small angles \(\alpha_{1,2}\) we can take the picture in Fig~\ref{fig:OrthoReconnected} and tilt the asymptotic plane \(z=w=0\) into \(z'=w'=0\), with the region around the origin remaining essentially the same but expanding linearly with time.

If we rewrite the perturbed orthogonal state [Eq~\eqref{eq:OrthoReconnected}] in double polar coordinates, we obtain
\begin{align}
    \psi_{\perp} = r_1r_2e^{i(\ta_1+\ta_2)} - \gamma e^{i\beta},
    \label{eq:PolarOrthoRec}
\end{align}
where \(\gamma e^{i\beta} = a + ib\). Note that in these variables the equations for the core surface become
\begin{align}
    \ta_1+\ta_2=\beta, \quad r_1r_2=\gamma.
    \label{eq:PolarOrthoRecCore}
\end{align}
These expressions are very simple and give an immediate interpretation of \(\gamma\) and \(\beta\), but are not as conducive to plotting and visualisation as Eqs~\eqref{eq:OrthoRecCore} and \eqref{eq:OrthoRecCoreCrossed}. However, we can use Eq~\eqref{eq:PolarOrthoRecCore} to find an expression of the minimum distance between the origin and the perturbed orthogonal core. Recall that this distance was zero for the unperturbed orthogonal state (both vortex planes passed through the origin), so this gives us a measure of the perturbation which we will use later numerically. Using \(r_1^2+r_2^2=r^2\) we can define \(r_1=r\sin u\) and \(r_2=r\cos u\), with \(u\in[0,\pi/2]\). Then, substituting this into Eq~\eqref{eq:PolarOrthoRecCore}, we obtain \(r^2 = 2\gamma/\sin2u\), such that the minimum value of \(r\) occurs at \(u=\pi/4\), and is given by
\begin{equation}
r_{\text{min}}=\sqrt{2\gamma}.
    \label{eq:rMinGamma}
\end{equation} Furthermore, \(r\) takes this value when \(r_1=r_2=\sqrt{\gamma}\).

\begin{figure}[!]
    \centering
    \includegraphics[width=0.48\textwidth]{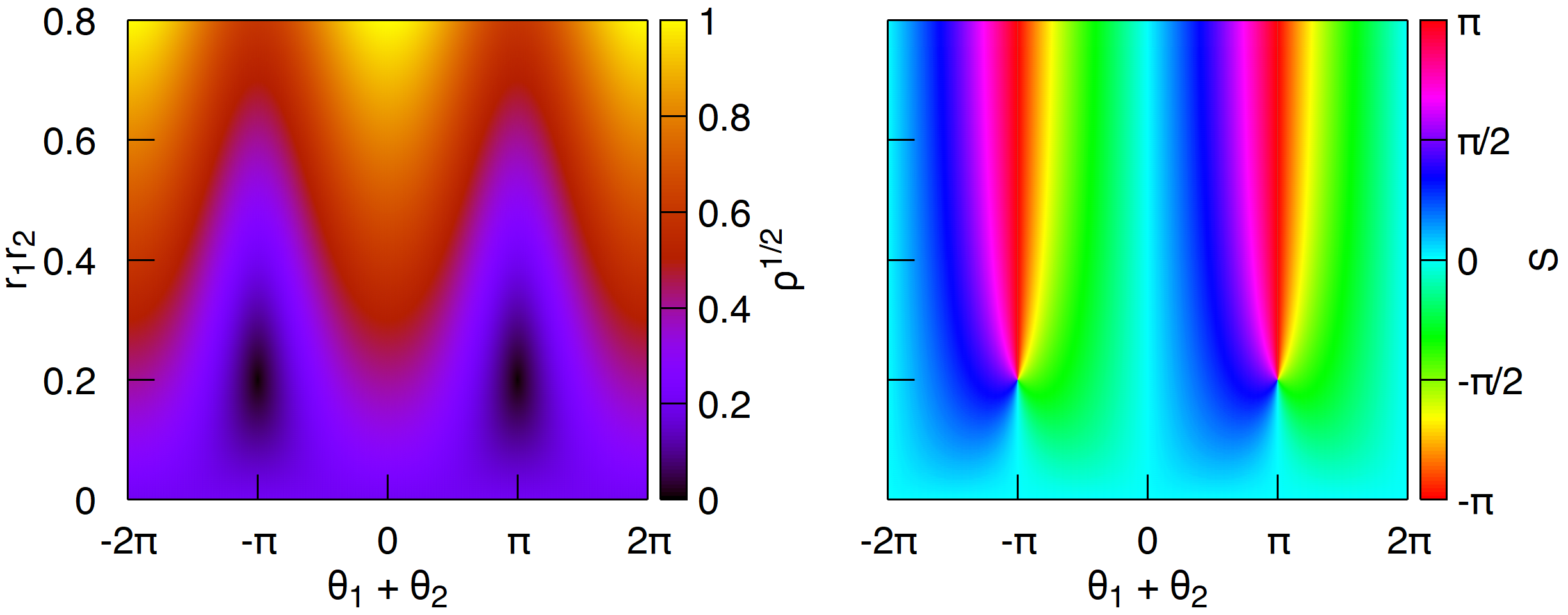}
    \caption{Density (Left) and Phase (Right) of the perturbed orthogonal state [Eq~\eqref{eq:PolarOrthoRec}] as a function of \(r_1r_2\) and \(\ta_1 + \ta_2\) as given by Eqs~\eqref{eq:OrthoRecDensity} and \eqref{eq:OrthoRecPhase}, respectively, for \(\gamma=0.2\) and \(\beta=\pi\). All quantities are dimensionless. Note that the zeroes of the density and corresponding branch points in the phase occur at \(r_1r_2=\gamma\) and \(\ta_1+\ta_2=\pm\beta\). The phase winds with \(\ta_1\) and \(\ta_2\) when \(r_1r_2>\gamma\), while it is roughly constant for \(r_1r_2<\gamma\).}
    \label{fig:AvoidedAnalytic}
\end{figure}

We can also use the polar form Eq~\eqref{eq:PolarOrthoRec} to plot the entire density and phase profiles by noting that this state is only a function of the variables \(r_1r_2\) and \(\ta_1+\ta_2\). Taking the modulus and argument of Eq~\eqref{eq:PolarOrthoRec} gives
\begin{align}
    &\rho = \left[r_1^2r_2^2 + 2\gamma r_1r_2\cos(\ta_1+\ta_2 - \beta) + \gamma^2\right]^{\frac{1}{2}},
    \label{eq:OrthoRecDensity} \\
    &\tan{S} = \frac{r_1r_2\sin(\ta_1+\ta_2-\beta)}{r_1r_2\cos(\ta_1+\ta_2-\beta) - \gamma},
    \label{eq:OrthoRecPhase}
\end{align}
which we have then plotted in the left and right panels, respectively, of Fig~\ref{fig:AvoidedAnalytic}, for \(\gamma=0.2,\) and \(\beta=\pi\). This plot clearly shows that the core --- visible as dark spots in the left panel and branch points in the right panel --- occurs at the values given in Eq~\eqref{eq:PolarOrthoRecCore}. Above \(r_1r_2=\gamma\) the phase winds once as either \(\ta_1\) or \(\ta_2\) makes a full circle, while below it the phase becomes approximately constant. Note that we have truncated the y-axis of this plot as we expect the linearised approach to only be meaningful within about a healing length of the core structure, where the healing length is given by $\xi=1$ in these dimensionless units. For the same reason, we only consider values of \(\gamma<1\); as we can identify $\gamma \propto \Delta t$, this is similar to the assumption that the evolution only describes short times.
To make this analogy clearer, we note that the skewed reconnecting state [Eq~\eqref{eq:reconnection}] at a fixed time step \(\Delta t\) has the same density and phase profiles as in Eqs~\eqref{eq:OrthoRecDensity} and \eqref{eq:OrthoRecPhase}, respectively, but with \(r_2\) and \(\ta_2\) replaced by \(r_2'\) and \(\ta_2'\) and with $\gamma=\Delta t (\sin \alpha_1 -\sigma \sin \alpha_2) $ and $\beta=-\pi/2$. 

In this simplified linear description the perturbation governing the avoided crossing grows linearly with time for short times; it would be very interesting in future work to extend this to the nonlinear regime using analogous methods to those of Nazarenko and West~\cite{nazarenko2003}, in order to discover the fate of these reconnected planes at later times. In particular, the question arises whether these curved vortex core structures can ever form a stationary state. As we shall see later, our numerical results show vortex core structures that are qualitatively similar to the skewed avoided crossing states we have considered here. However, these numerical vortex cores come from final states of the ITEM, which are numerical stationary states; also, these states have avoided crossing regions spanning several healing lengths, and so we do not expect these to be described by the linearised GPE [Eq~\eqref{eq:Schrodinger}].

Additionally, in this section we have assumed an infinite condensate, while our numerics uses a hardwall boundary. Once this boundary condition is imposed the off-axis nature of the avoided crossing introduces unavoidable image effects. This can be seen by evaluating the current of the reconnecting skewed state [Eq~\eqref{eq:reconnection}], given by \(\text{Im}\left( \psi^*\nabla\psi \right) = \rho\mathbf{v}\), and seeing that there is a radial component of the velocity at the boundary. Physically we require that the radial velocity at the boundary vanishes, otherwise the condensate would be expanding and the state would not be stationary. In fact, the 2D equivalent of Eq~\eqref{eq:OrthoReconnected}, that is \(\psi = x+iy - a - ib\), is precisely an off-axis point vortex located at the coordinates (\(a,b\)) rather than the origin, and is well known not to satisfy this boundary condition for the velocity field~\cite{Newton2001}. This condition is then usually enforced analytically with the method of images, but this is not straightforward given the complicated curved geometry of the vortex core. For this reason, analysis of these image effects is beyond the scope of this paper. 

Bearing these caveats in mind, in the next section we will develop a theory of superfluids doubly rotating with equal frequencies using intersecting non-orthogonal vortex planes as an ansatz for the ground state. Numerically, we will then observe stationary states with approximately the structure of this ansatz but with avoided crossings instead of intersections reminiscent of the reconnection physics that we have discussed in this section.

\section{Equal frequency double-rotations}
\label{sec:Equal}

In Ref.~\cite{mccanna2021}, we considered the case of a superfluid undergoing constant left isoclinic rotation in time, given by \(\Omega_{xy} = \Omega_{zw} \equiv \Omega\) in the lab (\(x,y,z,w\)) frame. We showed how this type of equal-frequency double rotation could energetically stabilise a configuration of two completely orthogonal vortex planes that intersect at a point, as reviewed in Sec~\ref{sec:4Dsuper}. Here we will consider this case again, but with a more generalised ansatz that includes the possibility of those vortex planes tilting away from the rotation planes and towards each other, in an anti-aligning sense, in order to benefit from attractive interaction energy at the expense of increased rotational energy~\cite{mccannaunequal}. We will use both analytics and numerics to investigate whether such a state can be energetically preferred to an orthogonal one.

\subsection{Analytics for a pair of tilted vortex planes under equal-frequency double rotation}

We shall begin by discussing an ansatz for a pair of vortex planes that are each tilted away from the planes of rotation, before using this ansatz to analytically compare the rotational and vortex-vortex interaction energies associated with such a structure. From this energetic balance, we shall analyse whether such a pair of tilted planes is expected to be favoured as compared to a pair of orthogonal planes under equal-frequency double rotation. Finally, in Sec.~\ref{sec:numerical}, we shall present and discuss corresponding numerical results. 

\subsubsection{Ansatz for a Pair of Tilted Anti-Aligning Planes}

We can write down a general ansatz for a pair of tilted vortex planes in the following form
\begin{align}
    \psi &= n^{\frac{1}{2}}e^{i\act_1}e^{i\grt_2},
    \label{eq:Equal:skewAnsatz}
\end{align}
where the acute (\(\acute{r}\)) and grave (\(\grave{r}\)) coordinate systems are each defined with respect to one of the tilted vortex planes, as will be specified in greater detail below. Here, we have also assumed that we can ignore the density depletion around the vortex core, and have instead approximated the density as a constant $n = N / V$, where $N$ is the number of particles and $V$ is the total volume. This assumption simplifies the energy calculations in the following section, and should be suitable for a large enough system size. 

We also note that the ansatz in Eq.~\eqref{eq:Equal:skewAnsatz} assumes that the vortex planes are flat and intersect at the origin. In particular this means we are not going to consider the conclusion from Sec~\ref{sec:Reconnection} that, in general, intersecting non-orthogonal vortex planes do not form a stationary state but will instead reconnect. This reconnection was found to cause the vortex surfaces to move outward near their intersection, such that they no longer intersect but instead form a single smoothly curved surface. This displacement of the cores will affect the velocity field directly, by moving the center of the circulating flow, and (if there is a hard-wall boundary) indirectly, by introducing image effects which we are as yet unable to model. Note, however, that the reconnected vortex cores (see Fig~\ref{fig:OrthoReconnected}) approach the original vortex planes for large values of \(r_1\) or \(r_2\). Therefore, if the typical size of the reconnection is small compared to the radius of the system, we may neglect these effects as they are only appreciable in a small region around the origin. 

To proceed, we would like to specify how the acute (\(\acute{r}\)) and grave (\(\grave{r}\)) coordinate systems are defined. To start, we note that, in general, we can write each in terms of the lab frame as \(\acbf{r} = \acute{M}\mathbf{r} \), and \(\grbf{r} = \grave{M}\mathbf{r}\), where \(\acute{M}\) and \(\grave{M}\) are arbitrary rotations. However, any element \(M\) of \(SO(4)\) can also be written as a product of some left isoclinic rotation \(M_L\) and some right isoclinic rotation \(M_R\), and that these always commute~\cite{mccannaunequal}. Therefore, we define \(\acute{M} = \acute{M}_L\acute{M}_R\) and \(\grave{M} = \grave{M}_L\grave{M}_R\). The relationship between the two tilted coordinate systems is then given by
\begin{align}
    \acbf{r} &= \acute{M}\grave{M}^{-1}\grbf{r}, \\
    &= (\acute{M}_L \grave{M}_L^{-1}) (\acute{M}_R \grave{M}_R^{-1}) \grbf{r}.
    \label{eq:AntiSkewIsoDecomp}
\end{align}
In order for the hydrodynamic vortex-vortex interaction between the two planes to be attractive, they should be anti-aligning~\cite{mccannaunequal}, as also mentioned above. Here this means that the transformation between their respective coordinate systems, (and hence the entire matrix product in Eq~\eqref{eq:AntiSkewIsoDecomp}) must be a left isoclinic rotation, see discussion in Ref.~\cite{mccannaunequal}. Now, the product of two left (right) isoclinic rotations is always a left (right) isoclinic rotation, and so the first factor in brackets in Eq~\eqref{eq:AntiSkewIsoDecomp} is a left isoclinic rotation while the second factor is a right one. Therefore for \(\acbf{r}\) and \(\grbf{r}\) to be related by a left isoclinic rotation we must reduce the second bracketed term to the identity matrix, by setting \(\acute{M}_R=\grave{M}_R \equiv M_R\). The definitions of each of the coordinates then become \(\acbf{r} = \acute{M}_L M_R \mathbf{r}\), and \(\grbf{r} = \grave{M}_L M_R \mathbf{r}\). 

To proceed, we note that, as commented in Sec~\ref{sec:4DRot}, an isoclinic double rotation does not have a unique pair of rotation planes, but instead an infinite set of them. Furthermore, it can be shown that the right isoclinic rotations are precisely the transformations between the rotation planes of a left isoclinic rotation and vice-versa~\cite{mccannaunequal}. This means that a spherically symmetric 4D system subjected to constant left isoclinic rotation in time, $M_L(t)$, has symmetry with respect to all right isoclinic rotations. In the case of a superfluid rotating in this way, this means there is a degenerate set of orthogonal vortex states corresponding to the set of orthogonal pairs of rotation planes of $M_L(t)$, as mentioned in Sec.~\ref{sec:4Dsuper}.
For our purposes here, this means that we can use the symmetry with respect to right isoclinic rotations to redefine \(\mathbf{r}\to M_R^{-1}\mathbf{r}\), absorbing \(M_R\) into the definition of the lab frame. The acute and grave coordinates are each then related to the lab frame, and to each other by a left isoclinic rotation ($\acute{M}_L \grave{M}_L^{-1}$). 
Using the relations reviewed in Sec~\ref{sec:4DRot}, we can then explicitly write down the definition of the tilted coordinates as follows
\begin{align}
    \begin{pmatrix}
        \acr_1e^{i\act_1} \\
        \acr_2e^{-i\act_2}
    \end{pmatrix}
    &= 
    \begin{pmatrix}
        \cos\eta_1 & e^{i\varphi_1}\sin\eta_1 \\
        -e^{-i\varphi_1}\sin\eta_1 & \cos\eta_1
    \end{pmatrix}
    \begin{pmatrix}
        \cpolar{1} \\
        \ccpolar{2}
    \end{pmatrix},
    \label{eq:Equal:acute}
    \\
    \begin{pmatrix}
        \grr_1e^{i\grt_1} \\
        \grr_2e^{-i\grt_2}
    \end{pmatrix}
    &= 
    \begin{pmatrix}
        \cos\eta_2 & e^{i\varphi_2}\sin\eta_2 \\
        -e^{-i\varphi_2}\sin\eta_2 & \cos\eta_2
    \end{pmatrix}
    \begin{pmatrix}
        \cpolar{1} \\
        \ccpolar{2}
    \end{pmatrix}. \quad
    \label{eq:Equal:grave}
\end{align}
The location of each vortex plane is then set by \(\acr_1 = 0\) and \(\grr_2 = 0\), respectively. The parameters \(\eta_{1,2}\) denote the angle between each plane and the \(x\en y\) (resp. \(z\en w\)) plane. The angles \(\varphi_{1,2}\) in Eq.~\eqref{eq:Equal:grave} then denote the direction of the tilts; however, 
it can be shown that the sum \(\varphi_1+\varphi_2\) can be chosen arbitrarily by a change of basis~\cite{mccannaunequal}, which allows us to set \(\varphi_1+\varphi_2=\pi\). Then the difference \(\varphi_1-\varphi_2\) controls the relative direction of the tilting of the two planes~\cite{mccannaunequal}; as we want the vortices to be tilted directly towards each other, we can set this difference to zero, such that \(\varphi_1 = 0\), \(\varphi_2=\pi\). Then we finally arrive at 
\begin{align}
    \acr_1e^{i\act_1} &= \cos\eta_1\cpolar{1} + \sin\eta_2\ccpolar{2}, 
    \label{eq:Equal:acute1}\\
    \grr_2e^{i\grt_2} &= \cos\eta_2\cpolar{2} + \sin\eta_2\ccpolar{1}.
    \label{eq:Equal:grave2}
\end{align}
We can also define the ``skewness" between the pair of planes as \(\eta\equiv \eta_1+\eta_2\); this is an angle which measures how far from being mutually orthogonal the vortex planes are, and is chosen such that the angle between the two planes is given by \(\pi/2-\eta\)~\cite{mccannaunequal}.  Note that as the vortices are indistinguishable, the acute and grave coordinates can be chosen such that the vortex determined by Eq.~\ref{eq:Equal:acute1} is closer to the \(x\en y\) plane than that from Eq.~\ref{eq:Equal:grave2}, which translates to the constraint~\cite{mccannaunequal}
\begin{align}
    \eta \equiv \eta_1 + \eta_2\leq \frac{\pi}{2} .
    \label{eq:EtaRestriction}
\end{align}
However, in reality this sum should be restricted to an even smaller value because our constant-density approximation will give an unphysical divergent vortex-vortex interaction energy as \(\eta\to\pi/2\) and the vortex cores approach each other. This divergence is important as we are considering attractive interaction, since it incorrectly implies the energy can decrease without bound. With this in mind we will now calculate the rotational and vortex-vortex interaction energies of this configuration.

\subsubsection{Balancing Rotational and Interaction Energies}

Firstly, we have the rotational energy, which will be decreased due to the vortex planes tilting off of the manifold of rotation planes of the condensate. This means that such a configuration is less energetically favourable from a rotational point-of-view than untilted vortex planes. The form of our ansatz [Eq~\eqref{eq:Equal:skewAnsatz}] allows us to write the rotational energy density as
\begin{align}
    \psi^*\odl\psi = n\Omega\left( \frac{\hat{L}_+e^{i\act_1}}{e^{i\act_1}} + \frac{\hat{L}_+e^{i\grt_2}}{e^{i\grt_2}} \right),
\end{align}
where \(\hat{L}_+ = \hat{L}_{xy} + \hat{L}_{zw}\) and where we have used that $\Omega_{xy}= \Omega_{zw} \equiv \Omega$ for equal-frequency double rotation. We can rewrite the terms inside the brackets in terms of the lab frame by substituting in Eqs~\eqref{eq:Equal:acute1} and \eqref{eq:Equal:grave2} and using that $\hat{L}_+ \equiv - i \hbar \partial_{\theta_1}- i \hbar \partial_{\theta_2}$. For the first term this reads
\begin{align}
    \frac{\hat{L}_+\acr_1e^{i\act_1}}{\acr_1e^{i\act_1}} &= \hbar \frac{\cos\eta_1\cpolar{1} - \sin\eta_1\ccpolar{2}}{\cos\eta_1\cpolar{1} + \sin\eta_1\ccpolar{2}}, 
\end{align}
where we have also divided through by $\acr_1$. By using the product rule and rearranging, we can then see that
\begin{align}
    \frac{\hat{L}_+e^{i\act_1}}{e^{i\act_1}} &= \hbar \left(1 - \frac{2 \tan\eta_1r_2 e^{-i(\ta_1+\ta_2)}}{r_1+\tan\eta_1r_2e^{-i(\ta_1+\ta_2)}} \right) - \frac{\hat{L}_+\acr_1}{\acr_1}. 
\end{align}
To obtain the total rotational energy, we need to integrate the LHS over the 4D superfluid. Following the method of Ref.~\cite{mccannaunequal}, this can be done using complex analysis to give
\begin{align}
    \dint_{B^4(R)} \diff^4 r \frac{\hat{L}_+ e^{i\act_1}}{e^{i\act_1}}
    &= \hbar \frac{\pi^2}{2}R^4 \left(1 - 2\sin^2\eta_1\right), \\
    &= \hbar V \cos2\eta_1,
\end{align}
where $B^4(R)$ is 4D hyperball of radius $R$, and $V = \pi^2 R^4/2$ is its volume. Similarly, the other term above integrates to \( \hbar V\cos2\eta_2\) by symmetry. The rotational energy reduction is then
\begin{align}
    E_{\text{rot}} &= \int\psi^*\odl\psi\diff^4 r \\
    &= N \hbar \Omega \left( \cos2\eta_1 + \cos2\eta_2 \right).
\end{align}

Secondly, we need to consider the hydrodynamic kinetic energy, which can be written as
\begin{align}
    \frac{1}{2}\int\rho {\bf v}^2 \diff^4 r = \frac{1}{2}\int\rho \left({\bf v}_1^2 + {\bf v}_2^2\right) \diff^4 r + \int\rho \mathbf{v}_1\cdot\mathbf{v}_2 \diff^4 r.
\end{align}
Here, the first term on the RHS is the individual hydrodynamic cost of each vortex, while the second term is the vortex-vortex interaction energy~\cite{mccanna2021, mccannaunequal}. Note that in our ansatz [Eq~\eqref{eq:Equal:skewAnsatz}] we assumed that the density was constant (i.e. ignoring the vortex core), in which case, the first term on the RHS will diverge. However, for the purposes of this calculation, we are only interested in how the energy changes as the vortex planes tilt. As this first term does not vary with the orientation of the planes due to the boundary's spherical symmetry, we will ignore this term hereafter. We also note that by making a constant density approximation, we are ignoring energy contributions from quantum pressure, and bosonic interactions, as is also commonly done in 2D superfluids~\cite{pethick2002, pitaevskii2003}.

The above considerations leave us with only the vortex-vortex interaction term, which clearly vanishes for an orthogonal pair of vortex planes where $\mathbf{v}_1\cdot\mathbf{v}_2 =0 $. For tilted planes, this interaction energy is generally non-vanishing and can be derived as~\cite{mccannaunequal}
\begin{align}
    E_{\text{vv}} = 4\mu N \frac{\xi^2}{R^2}\ln\cos(\eta_1+\eta_2),
\end{align}
which is negative and hence attractive. It also diverges as $\eta \equiv \eta_1+\eta_2 \rightarrow \pi/2$, corresponding to the limit that the two vortex planes overlap, in which case the constant density approximation will break down, as discussed above. 

\begin{figure*}[t!]
    \centering
    \includegraphics[width=\textwidth]{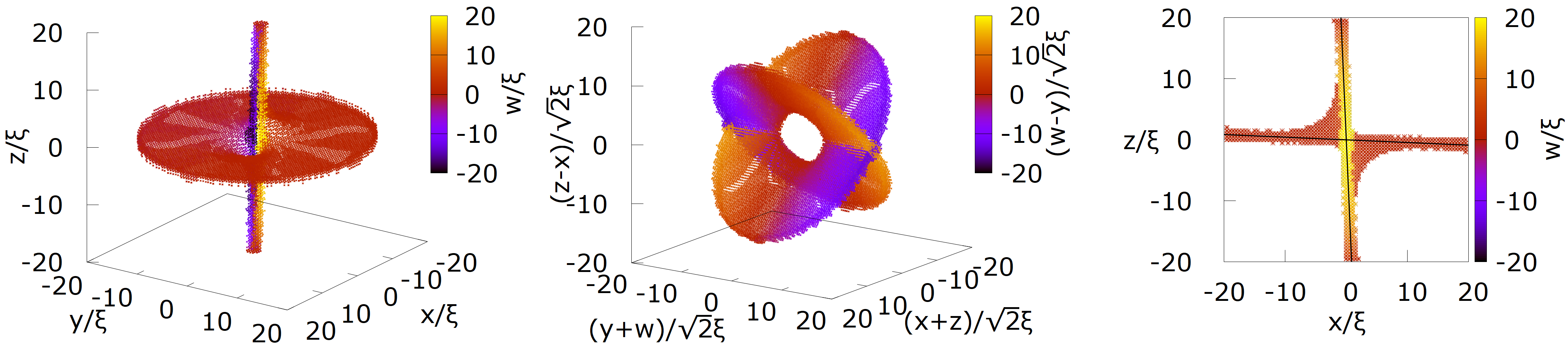}
    \caption{Numerical anti-aligning non-orthogonal vortex core stationary state for the parameters \(\Omega=0.75\Omega_c\), \(\Delta x = 0.5\xi\), and \(R\approx20.6\xi\). The first panel shows the core points in \((x,y,z,w)\) space, with the \(w\) value represented as colour. This view shows the small tilt angle clearly, but obscures the avoided crossing region centred around the origin as one of the planes is collapsed towards the vertical axis. The second panel shows the same data after double rotation given in Eq~\eqref{eq:CrossedRotation} (see the axes and colourbar labels for details), which shows the avoided crossing region clearly. The third panel shows a side-on view, along with a by-eye estimate of the tilt angle with the lines \(z=-x\tan(2.5\degree)\) and \(z=-x/\tan(2.5\degree)\). Note that at this frequency both this state and the orthogonal state are higher in energy than the state with no vortices.}
    \label{fig:F0.75AntiSkew}
\end{figure*}

Now, to find the most energetically favourable state, we can define a dimensionless energy density relative to \(E_{\text{rot}}^{\perp}=2N \hbar \Omega\) (corresponding physically to the rotational energy of two orthogonal vortex planes under equal-frequency double-rotation) as 
\begin{align}
    \varepsilon = \frac{R^2}{4\xi^2\mu N}(-E_{\text{rot}} + E_{\text{rot}}^{\perp} + E_{\text{vv}}).
\end{align}
We also introduce a dimensionless frequency \(\omega = R^2\hbar\Omega/2\xi^2\mu\), such that then
\begin{align}
    \varepsilon = \omega \left[ 1 - \frac{1}{2}\left(\cos2\eta_1 + \cos2\eta_2\right) \right] + \ln\cos(\eta_1+\eta_2). \
    \label{eq:AntiTiltEnergy}
\end{align}
Taking derivatives of this energy we find that
\begin{align}
    \frac{\partial\varepsilon}{\partial\eta_1} &= \omega\sin2\eta_1 - \tan(\eta_1+\eta_2) = 0, \\
    \frac{\partial\varepsilon}{\partial\eta_2} &= \omega\sin2\eta_2 - \tan(\eta_1+\eta_2) = 0.
\end{align}
This implies that \(\sin2\eta_1=\sin2\eta_2\), such that either \(\eta_1=\eta_2\) or \(\eta_1 = \pi/2 - \eta_2\). However, we can rule out the latter solution as this results in the vortices coinciding, which is a limit that our current approximations break down in. We therefore take \(\eta_1=\eta_2\equiv \eta/2\) and proceed. Both equations above then become
\begin{align}
    \omega\sin\eta = \tan\eta,
\end{align}
which implies either \(\sin\eta=0\) or \(\cos\eta=1/\omega\). The former condition gives \(\eta=0\) --- the orthogonal state --- while the latter looks like a promising candidate for a skewed state that is lower in energy, provided \(\omega>1\). However, if we look at the energy of this state we find \(\varepsilon = \omega - 1 - \ln\omega\), which is never negative. This state is therefore always higher energy than the orthogonal state. A more detailed analysis reveals that it is a saddle point in the \(\eta_{1,2}\) energy landscape. Therefore, this theory predicts that for \(\omega>1\) the orthogonal state at \(\eta=0\) is a local minimum and there is an energy barrier for the states to tilt away from this. For \(\omega\leq1\), both the minimum and saddle point disappear, and the predicted energy decreases monotonically with \(\eta\).

In the latter case the orthogonal state would be unstable to this form of anti-aligned tilting even to the limit \(\eta_1=\eta_2\to\pi/4\) where the vortex planes lie on top of each other with opposite winding. Eq~\eqref{eq:AntiTiltEnergy} predicts that \(\varepsilon\to-\infty\) in this limit, but this is only because our constant density approximation for the interaction energy fails as the vortex cores increasingly overlap. In reality vortices that coincide in opposite senses annihilate each other, so we can interpret the \(\omega<1\) regime as suggesting that the orthogonal state is unstable below a certain threshold frequency. Note that in units of the critical frequency, \(\Omega_c\), the dimensionless frequency is given by \(\omega = \ln(2.07R/\xi)\Omega/\Omega_c\), so this threshold frequency is given in terms of the critical frequency as \(\Omega_{stab} = \Omega_c/\ln(2.07R/\xi)\). This will always be smaller than \(\Omega_c\) for \(R>1.32\xi\), so this stability threshold does not alter the critical frequency. Investigating this predicted stability threshold is beyond the scope of this paper but would be an interesting topic for further work.

\subsection{Numerical Results}
\label{sec:numerical}

Now we numerically test the analytical prediction that the orthogonal state is lower energy than any anti-aligning state for \(\omega>1\). To do so, we use the imaginary time evolution method (ITEM) as described in Appendix~\ref{app:methods} for initial states with a phase profile corresponding to the anti-aligning ansatz [Eq~\eqref{eq:Equal:skewAnsatz}] with a uniform density away from the boundary, and added noise. By symmetry we assume that \(\eta_1=\eta_2\), although the actual value of this angle must be chosen arbitrarily, as we have no predicted state to inform us. For the results presented in this section we chose an initial tilt angle of \(\eta_1=\eta_2=5\degree\).

We ran the ITEM with the parameters \(\Omega = 0.75\Omega_c\), \(\Delta x=0.5\xi\), and \(R\approx20.6\xi\) on our initial state and then calculated the energy and vortex core points from the final state. The resulting vortex core is plotted in Fig~\ref{fig:F0.75AntiSkew}, and looks like a pair of slightly skewed planes at large distances from the origin. The third panel shows the core side-on, ignoring the y coordinates, and shows lines plotted on top that give an estimate of the tilt angles by eye as \(\eta_1=\eta_2\approx2.5\degree\), such that the state has untilted slightly from our initial phase ansatz in a symmetric manner. This result suggests that the theory above is not a bad approximation: the final state that we have is close to a pair of orthogonal planes --- even closer in angle than our initial phase profile.

Despite this, there is an avoided crossing region near the origin, out to radii of several healing lengths, which is qualitatively similar to the core structure derived in Sec~\ref{sec:Reconnection} by considering the linearised dynamics of intersecting vortex planes. In particular, the first panel of Fig~\ref{fig:F0.75AntiSkew} shows a core geometry qualitatively similar to that of Fig~\ref{fig:OrthoReconnected}, with the avoided crossing obscured by the fact that much of the core is collapsing toward the vertical axis. To make the avoided crossing more visible, we rotated the coordinates according to the same double rotation [Eq~\eqref{eq:CrossedRotation}] that was used to make the second panel of Fig~\ref{fig:OrthoReconnected} and plotted the data against this new basis in the second panel. Again, the resulting plot looks similar to the orthogonal perturbed state, except that the avoided crossing region is much larger than that allowed by the linearised analysis in Sec.~\ref{sec:Reconnection}. Another caveat we must make when drawing this analogy is that Fig.~\ref{fig:F0.75AntiSkew} shows the vortex core of a numerical stationary state, whereas Sec.~\ref{sec:Reconnection} dealt with dynamically evolving states.

We also investigate the frequency dependence of this physics by using the final state above as the initial state for another run of the ITEM with \(\Omega=0.80\Omega_c\), and then iterating this at regular frequency intervals up to \(\Omega=1.5\Omega_c\), outputting the energy and vortex core for each state. Finally, this entire loop was repeated for different values of the spatial resolution --- and hence the system radius --- from \(\Delta x=0.5\xi\), which gives \(R\approx20.6\xi\), down to \(\Delta x=0.25\xi\), which gives \(R\approx10.3\xi\), in steps of \(0.05\xi\). In order to find out whether these anti-aligning states are energetically favoured, we also found the orthogonal state for each of these system sizes and computed its energy for each of these frequencies. We find that the energy is almost always slightly lower than that of the orthogonal state, although this energy difference is very small and possibly dominated by numerical error. Higher accuracy simulations will be needed to investigate the difference between these states.

{
}

\begin{figure}
    \centering
    \includegraphics[width=\linewidth]{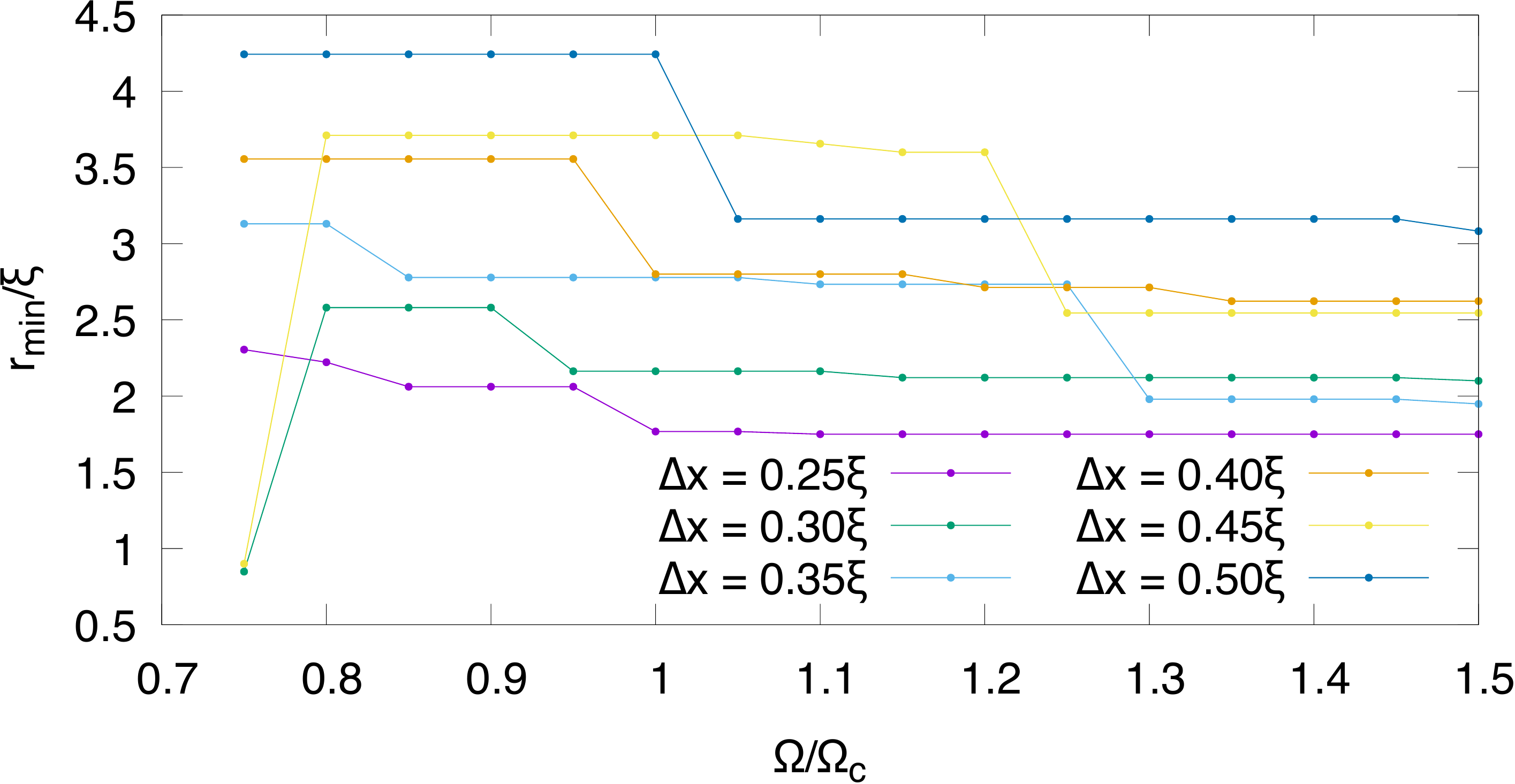}
    \caption{The minimum distance \(r_{\text{min}}\) from the vortex core to the origin for the states described in this section, as a function of \(\Omega\) and \(\Delta x\), giving a measure of the size of the avoided crossing region. The data is not very smooth, suggesting there are multiple energetically-close metastable branches being explored. There may also be some sampling error due to the discretisation of the Cartesian grid. Nevertheless, the general trends are that \(r_{\text{min}}\) decreases with \(\Omega\) and increases with \(\Delta x\) (and hence system size). The first trend makes sense if the avoided crossing is reducing the state's angular momentum, while the second trend does not have a clear explanation. The two outlier points at \(\Omega=0.75\Omega_c\) (with \(r_{min} < 1\)) are cases where the final state has vortex cores skewed according to the initial phase profile tilt angle of \(5\degree\), likely due to insufficient accuracy given how small the energy differences are.
    }
    \label{fig:AntiTiltDist}
\end{figure}

\begin{figure*}
    \centering
    \includegraphics[width=\textwidth]{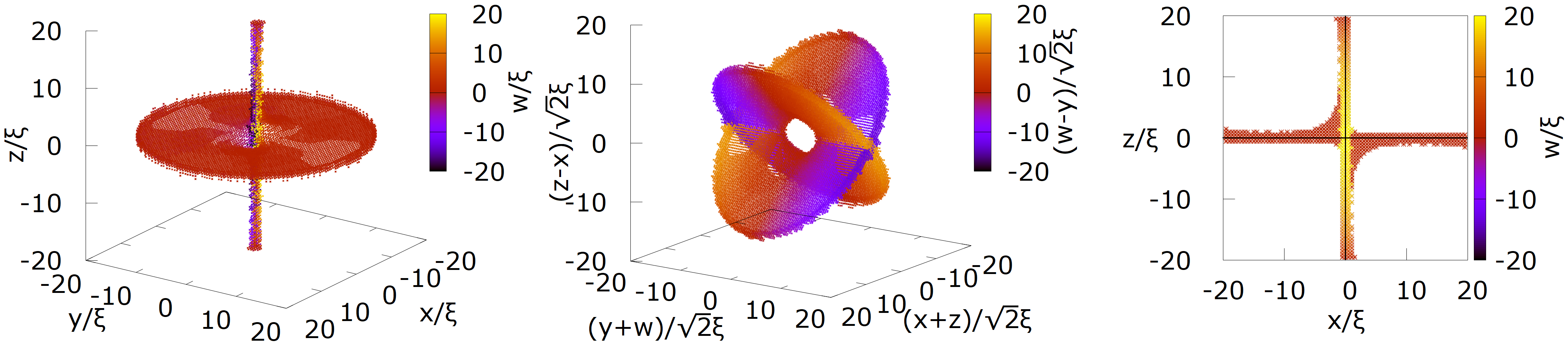}
    \caption{As Fig~\ref{fig:F0.75AntiSkew} but at a higher frequency of \(\Omega=1.5\Omega_c\) at which the vortex states are energetically favoured with respect to the no-vortex state. Compared to Fig~\ref{fig:F0.75AntiSkew} this state shows a reduction in both the tilt angle and the size of the avoided crossing. The by-eye estimate in the third panel shows a tilt angle of \(\approx0\degree\), that is, there is no discernible tilt. This can be intuitively understood as the higher frequency causes the negative rotational energy term to have a higher weighting relative to other terms in the energy. Consequently a higher angular momentum is favoured.}
    \label{fig:F1.50AntiSkew}
\end{figure*}

We can also estimate the size of the avoided crossing by calculating \(r_{\text{min}}\) (see Ref.~\cite{mccannaunequal}), the minimum distance between the vortex core and the origin. This is plotted in Fig.~\ref{fig:AntiTiltDist} for each of these states as a function of both \(\Omega\) and \(\Delta x\). Here we see that \(r_{\text{min}}>\xi\), such that these states lie outside the linearised GPE regime considered in Sec.~\ref{sec:Reconnection}. Furthermore, we see that, in general, the avoided crossing decreases in size with \(\Omega\), and increases with \(\Delta x\) (and hence system size). The former trend supports our intuition that the avoided crossing reduces the angular momentum, since higher frequencies favour higher angular momentum. The latter trend perhaps suggests that the attractive vortex-vortex interaction is not playing a large role in determining the avoided crossing size, since (for fixed particle number) the energy of this interaction decreases with increasing system size while \(r_{\text{min}}\) increases.

We can also see this behaviour by following the vortex core in Fig.~\ref{fig:F0.75AntiSkew} as we increase \(\Omega\) up to \(1.5\Omega_c\). Then we obtain a final state with the core structure shown in Fig.~\ref{fig:F1.50AntiSkew}. These plots show clearly, in the second panel, the reduction in size of the avoided crossing but they also show that this state is effectively no longer skewed. Looking at the third panel, we have superimposed the lines \(x=0\) and \(z=0\) and can see that these run parallel with the data points. This means that this state is effectively the orthogonal state with an avoided crossing (similar to Eq.~\eqref{eq:OrthoReconnected}), which is likely due to higher angular momentum being favoured at these higher frequencies. 
For a more detailed look at one of these avoided orthogonal states, including cuts of the density and phase profiles, at higher spatial resolution see Appendix~\ref{app:numerical}.

It is not clear whether both features of these states --- the skewness of the planes and the avoided crossing --- are important in lowering the energy. For this reason we have also performed the ITEM on initial states with a phase profile given by that of the orthogonal states plus a perturbation around the intersection point (as in Eq.~\eqref{eq:PolarOrthoRec} but with an unconstrained perturbation size). The final states of these numerical tests exhibit the same skewed and avoided crossing core structures as before, with the same trends in these features as the frequency and system size vary, indicating that both of these features are important. Note that if we set the perturbation to zero in the initial state the final state we obtain is the intersecting orthogonal state we previously studied~\cite{mccanna2021}.

\begin{figure}
    \centering
    \includegraphics[width=0.95\linewidth]{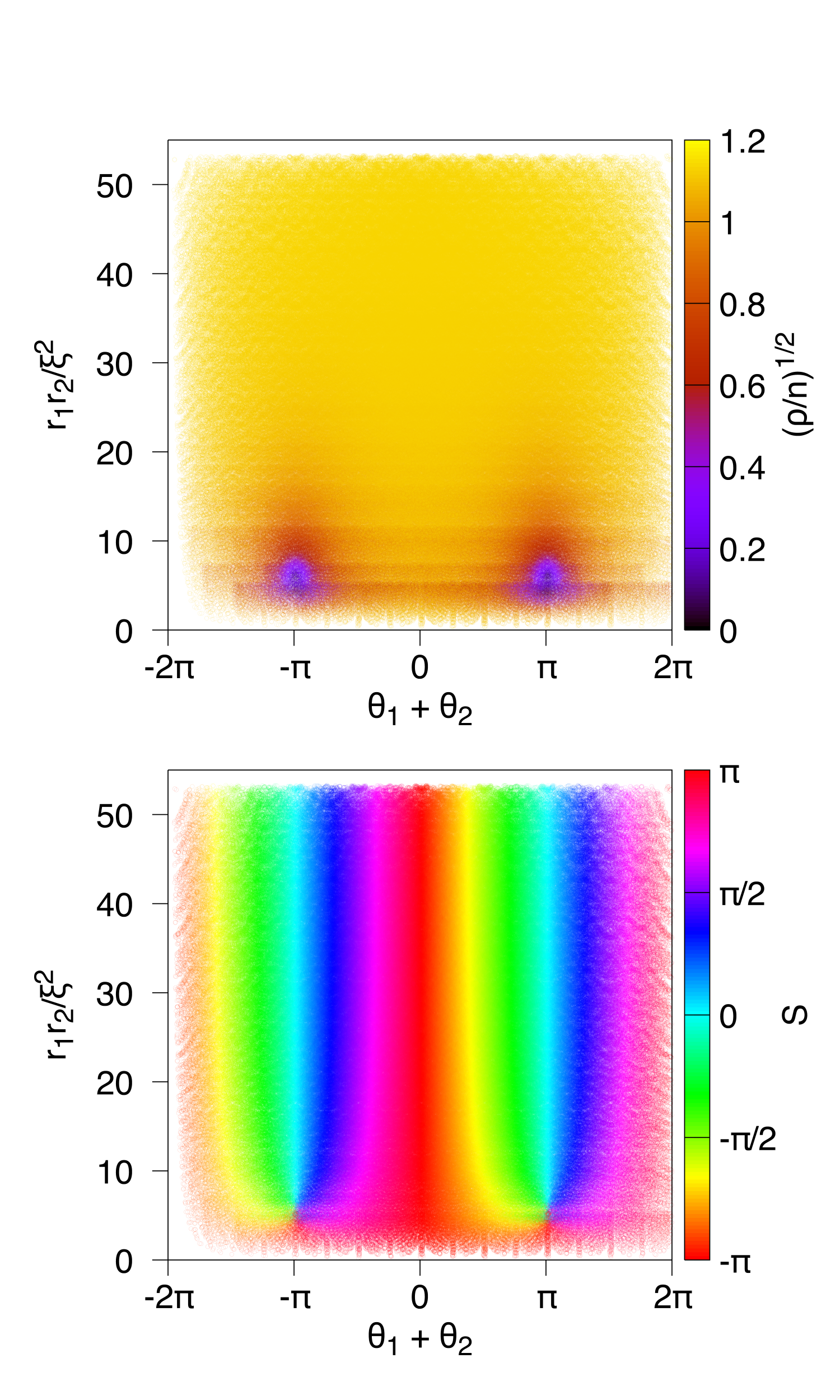}
    \caption{Density (top) and phase (bottom) profiles of the state whose core is shown in Fig.~\ref{fig:F1.50AntiSkew}. We have plotted all data points, up to a distance of roughly \(5\xi\) from the origin, according to their value of \(r_1r_2\) and \(\ta_1+\ta_2\), in order to compare to the analytic orthogonal perturbed state [Fig.~\ref{fig:AvoidedAnalytic}] which showed a similar core structure. In order to show points that are stacked below other points, we have added small random numbers to the \(x\) and \(y\) coordinates of each point and plotted them as open circles. This noise is sampled from a flat distribution ranging from \([0,0.025\pi]\) in the \(x\)-direction and \([0,0.25]\) in the \(y\)-direction. While neither profile is single-valued in terms of these variables, the overall structure --- particularly of the phase --- is very similar to that of Fig.~\ref{fig:AvoidedAnalytic}.}
    \label{fig:AntiTiltAvoided}
\end{figure}

Finally, we will investigate the analogy between these anti-aligning states and the linearised reconnection dynamics of Sec.~\ref{sec:Reconnection}. To do this we have used the final state at \(\Omega=1.5\Omega_c\) and \(\Delta x=0.25\xi\), which has an avoided crossing but no discernible skewness (as in Fig.~\ref{fig:F1.50AntiSkew}), so the core structure is very similar to that of analytic perturbed orthogonal state [Eq~\eqref{eq:PolarOrthoRec}]. Then, in Fig.~\ref{fig:AntiTiltAvoided} we have plotted all of the numerical data points up to a radius of roughly \(5\xi\) according to their value of \(r_1r_2\) and \(\ta_1+\ta_2\). These points are then coloured according to their value of \(\sqrt{\rho}\) in the left panel, and \(S\) in the right panel. This was inspired by Fig.~\ref{fig:AvoidedAnalytic} in Sec.~\ref{sec:Reconnection}, which shows the corresponding figure for the analytic perturbed orthogonal state, which depended only on \(r_1r_2\) and \(\ta_1+\ta_2\). Note that we have also added small random noise to the \(x\) and y \(coordinates\) of each point, to prevent points with the same value of \(r_1r_2\) and \(\ta_1+\ta_2\) from being perfectly stacked on top of each other, and hence not visible. 

In the case of the numerical data in Fig.~\ref{fig:AntiTiltAvoided}, we can see that the density and phase plots are not single-valued in terms of these variables, as there are regions with different colours stacked on top of each other. For this reason we have plotted the points as open circles so that they don't occlude each other as much. This multivaluedness is not surprising, as the numerical state is in the nonlinear GPE regime. Nevertheless, we do find many similarities between the overall structure of this plot and Fig.~\ref{fig:AvoidedAnalytic}. Firstly, the core (seen as dark spots in the density and branch points in the phase) is centred around a constant value of \(r_1r_2=\gamma\xi^2\) with \(\gamma\approx 4.5\), and \(\ta_1+\ta_2=\pm\pi\). This was a feature of the perturbed orthogonal state, although the corresponding value of \(\gamma\) was constrained to be small in this linearised case. Secondly, the phase profile appears to be very close to a single-valued function of these variables, with essentially the same behaviour as the analytic phase profile of the perturbed orthogonal state. In particular, we have that for \(r_1r_2<\gamma\) the phase is roughly constant, while for \(r_1r_2>\gamma\) the phase winds once as either \(\ta_1\) or \(\ta_2\) makes a full circle, which is exactly what we see in Fig.~\ref{fig:AvoidedAnalytic}. This suggests that there may be a similar analytic description for these numerical states --- or at least their phase profile.

\section{Conclusions}
\label{sec:concl}

In this paper, we have explored stationary states of the 4D GPE under equal-frequency double rotation, showing that these can have vortex cores formed of skew planes and curved surfaces. This work extends previous studies in Refs.~\cite{mccanna2021} and~\cite{mccannaunequal}, which focused on completely orthogonal and rigid vortex planes, and on unequal-frequency double rotation respectively. Interestingly, none of these states have a direct analogue in 2D and 3D rotating superfluid systems, showing that there is much rich vortex physics to be explored in higher spatial dimensions.

In more detail, in Sec.~\ref{sec:Reconnection} we used linearised GPE dynamics to show that the intersection point between a non-orthogonal pair of vortex planes is not stable in general, but undergoes a form of reconnection. In contrast to the reconnection of extended vortex lines in 3D, the core of the non-orthogonal vortices in 4D forms a single connected object at all times, with the intersection point replaced by an avoided crossing that expands linearly with a speed determined by the tilt angles. However, as we saw later in numerics, very similar core structures can become stable, suggesting that the GPE nonlinearity may limit the predicted expansion at a certain size. Investigating this and other potential causes for the stability of avoided crossings is an interesting possible avenue for future research.

Next, in Sec.~\ref{sec:Equal} we developed an analytic model for a superfluid under isoclinic, i.e. equal frequency, double rotation, to see whether a pair of planes skewed in an anti-aligning sense could have lower energy than an orthogonal state by benefiting from attractive vortex-vortex interactions at the expense of rotational energy. With this analysis, we found that skewed states are unlikely to be both stable and lower energy than the orthogonal state. In particular, above a threshold frequency \(\Omega_{\text{stab}}\) the orthogonal state was shown to be a local minimum, with an energy barrier to any anti-aligned configuration. The frequency \(\Omega_{\text{stab}}\) was found to be related to, but less than, the critical frequency \(\Omega_c\), such that the orthogonal state remains stable above \(\Omega_c\). However, this result does have implications for the metastability of the orthogonal state, since below \(\Omega_{\text{stab}}\) the vortex planes can continuously lower their energy by tilting toward each other in an anti-aligning sense, until eventually they come together and annihilate. Numerically testing this prediction is another possible avenue of research, but for now we have focused on the region above \(\Omega_{\text{stab}}\). There we have found that states with a small anti-aligning skewness and a sizeable avoided crossing can be (to the numerical accuracy we are working at) essentially degenerate with the orthogonal intersecting state. Furthermore, the skewness in the final states decreases with frequency and appears to vanish, while the avoided crossing size decreases but not to zero. This suggests that avoided orthogonal states may be almost degenerate with the intersecting orthogonal states even at higher frequencies, although higher accuracy is needed to confirm or refute this.

In the future, it will be important to also look beyond the 4D GPE studied here in order to consider more realistic experimental models~\cite{mccanna2021, mccannaunequal}. Interest in 4D systems has been sparked by developments based on topological pumping~\cite{Thouless1983,verbin2013,Kraus2012Adiabatic,Kraus2012Fibonacci,Kraus2013,verbin2015,Lohse2016,Nakajima2016,Lohse2018,Zilberberg2018,ChengW2021,Chen2021}, ``synthetic dimensions"~\cite{Schreiber2010,Regensburger2011,Boada2012,Celi2014,Stuhl2015,Mancini2015,Luo2015,Gadway2015, wang2015,Livi2016,Meier2016,Ozawa2016,Yuan2016,Kolkowitz2017,Cardano2017,Ozawa2017,Wimmer2017,Martin2017,Signoles2017, Sundar2018,Wang2018,Chen2018,Baum2018,Peng2018,salerno2019quantized, viebahn2019matter,barbiero2019bose, Price2020,chalopin2020exploring,ChengD2021,Kanungo2022, ozawa2017synthetic, lustig2019photonic, Yuan2018,Yuan2019, yuan2020creating, Dutt2020, baum2018setting, chen2018experimental, cai2019experimental, wimmer2021superfluidity, price2019synthetic,crowley2019half,boyers2020exploring,lienhard2020realization, kang2020creutz,balvcytis2021synthetic,chen2021real,oliver2021bloch,englebert2021bloch}, artificial parameter spaces~\cite{sugawa2018second,lu2018topological, kolodrubetz2016measuring, wang2020exceptional,zhu2020four,palumbo2018revealing, chen2022synthetic} and the connectivity of classical electrical circuits~\cite{wang2020circuit,Price2018, yu2019genuine,li2019emergence,ezawa2019electric} amongst other approaches. In particular, in a ``synthetic dimension", a set of degrees of freedom are externally coupled together and then re-interpreted as lattice sites along an extra spatial dimension~\cite{Boada2012}, which may open up the prospect of experimentally exploring 4D superfluids in the future. However, as discussed in Ref.~\cite{mccannaunequal}, such experiments will likely have various attributes, such as discrete lattices, broken \(SO(4)\) rotational symmetry and unusual interaction terms, which are not present in the 4D GPE, and which will therefore require further modelling.  

It will also be interesting in the further work to look for other types of possible 4D topological structures, such as closed vortex surfaces that generalise vortex loops (including links and knots)~\cite{Proment2012,Proment2014,Scheeler2014,villois2017universal,villois2020,Proment2020}, or stationary states at even higher rotation frequencies, where it may be favourable to have even richer curved vortex surfaces~\cite{mccannaunequal}. In the longer-term, our work can be extended to consider other order parameters, such as those of spinor condensates which are known to host non-Abelian vortices in 3D~\cite{Kawaguchi,Machon}, or to eventually move towards the strongly-interacting fractional quantum Hall regime~\cite{Zhang2001}.\\

{\it Acknowledgements:} We thank Tomoki Ozawa, Mike Gunn, Iacopo Carusotto, Mark Dennis, Davide Proment and Russell Bisset for helpful discussions. 
This work is supported by the Royal Society via grants UF160112, RGF\textbackslash{}EA\textbackslash{}180121 and RGF\textbackslash{}R1\textbackslash{}180071 and by the Engineering and Physical Sciences Research Council [grant number EP/W016141/1].

\appendix

\begin{figure*}[t!]
    \centering
    \includegraphics[width=0.9\textwidth]{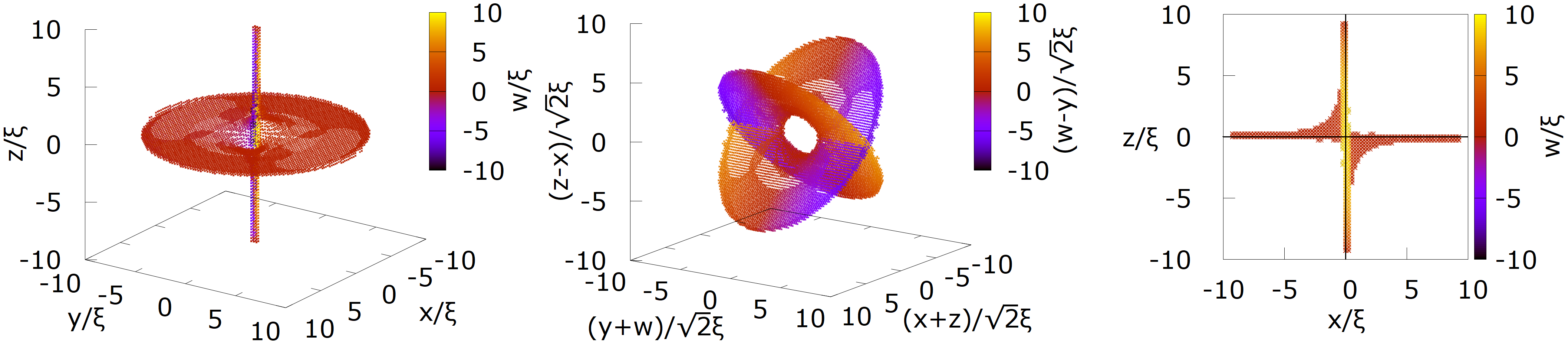}
    \caption{As Fig~\ref{fig:F1.50AntiSkew} but at a spatial step of \(\Delta x = 0.25\xi\), corresponding to a radius of \(R\approx10.3\xi\). Again, we have an avoided crossing but not visible skewness, with the lines \(x=0\) and \(z=0\) plotted on the figure as guides to the eye to show this. While it is not clear from comparing this figure to Fig~\ref{fig:F1.50AntiSkew}, the avoided crossing region is in fact smaller in this smaller system [c.f. Fig~\ref{fig:AntiTiltDist}]}
    \label{fig:HiResF1.50AntiSkew}
\end{figure*}

\begin{figure*}[t!]
    \centering
    \begin{minipage}{0.45\textwidth}
        \includegraphics[width=\linewidth]{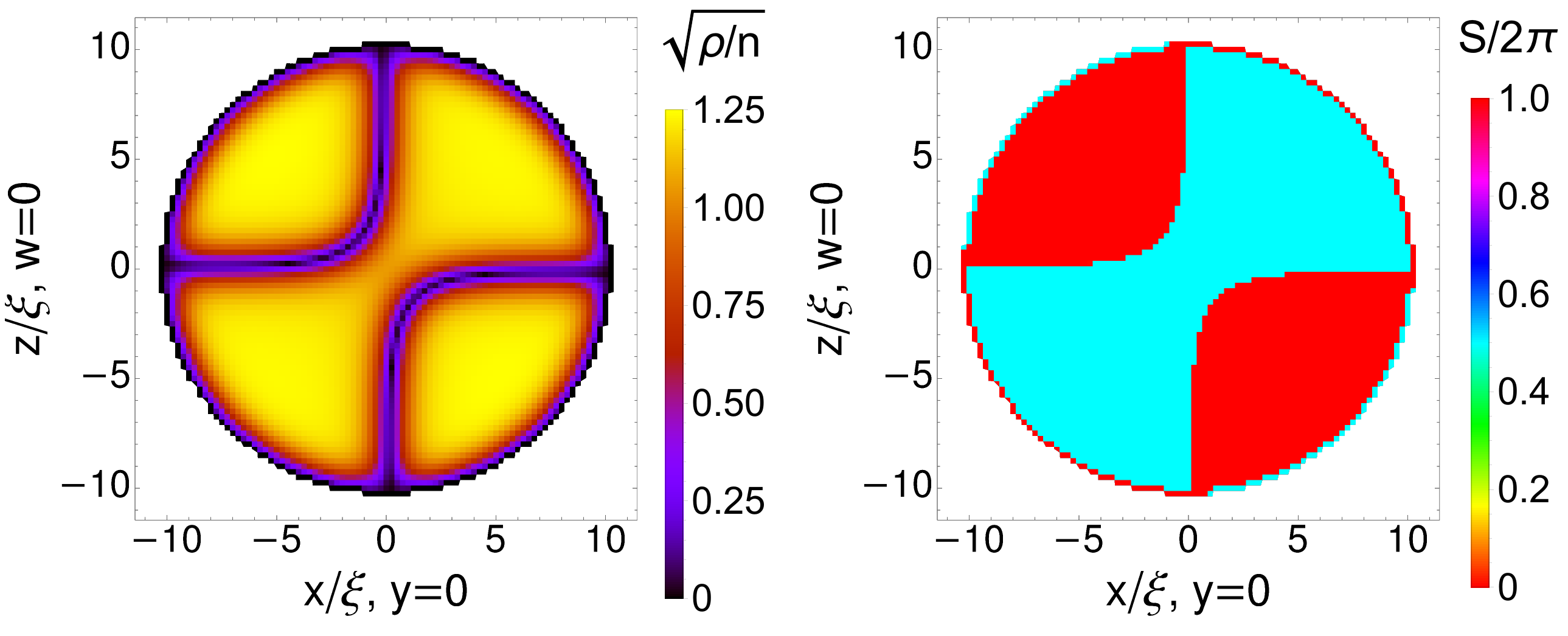}
    \end{minipage}
    \begin{minipage}{0.45\textwidth}
        \includegraphics[width=\linewidth]{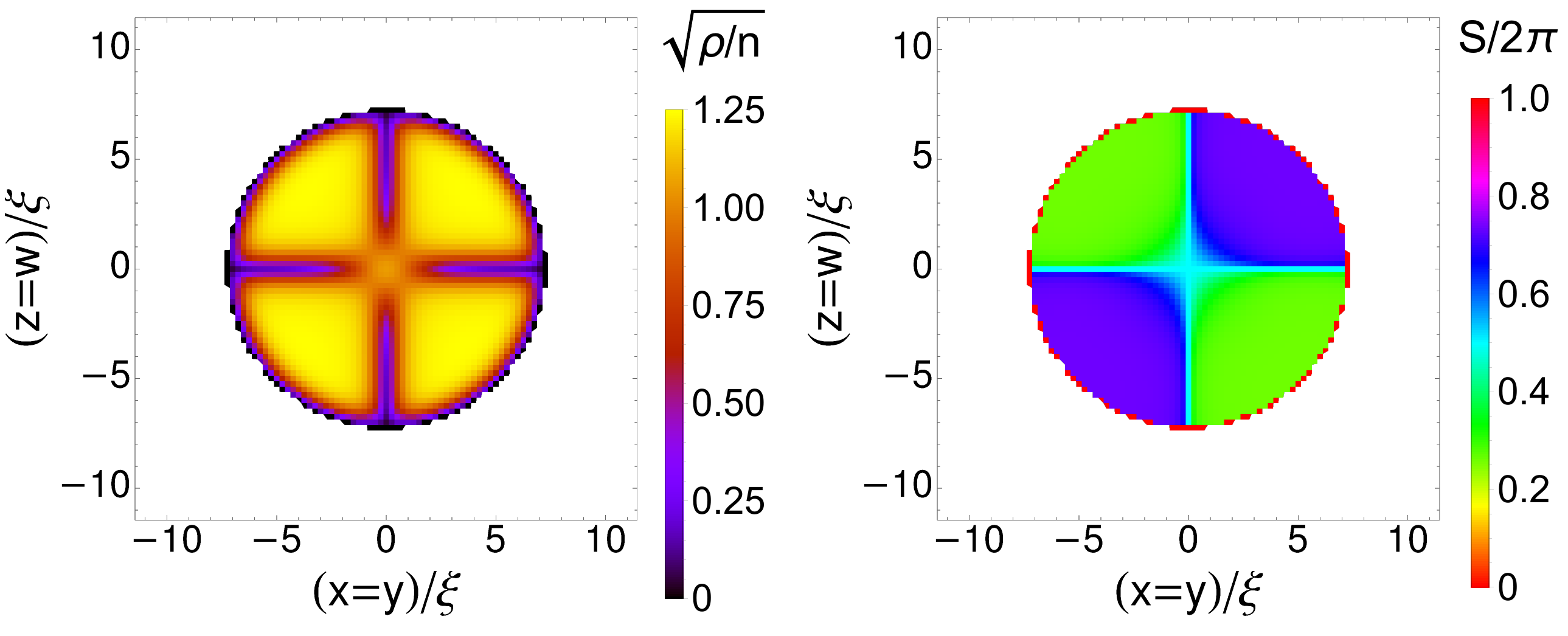}
    \end{minipage}
    \begin{minipage}{0.45\textwidth}
        \includegraphics[width=\linewidth]{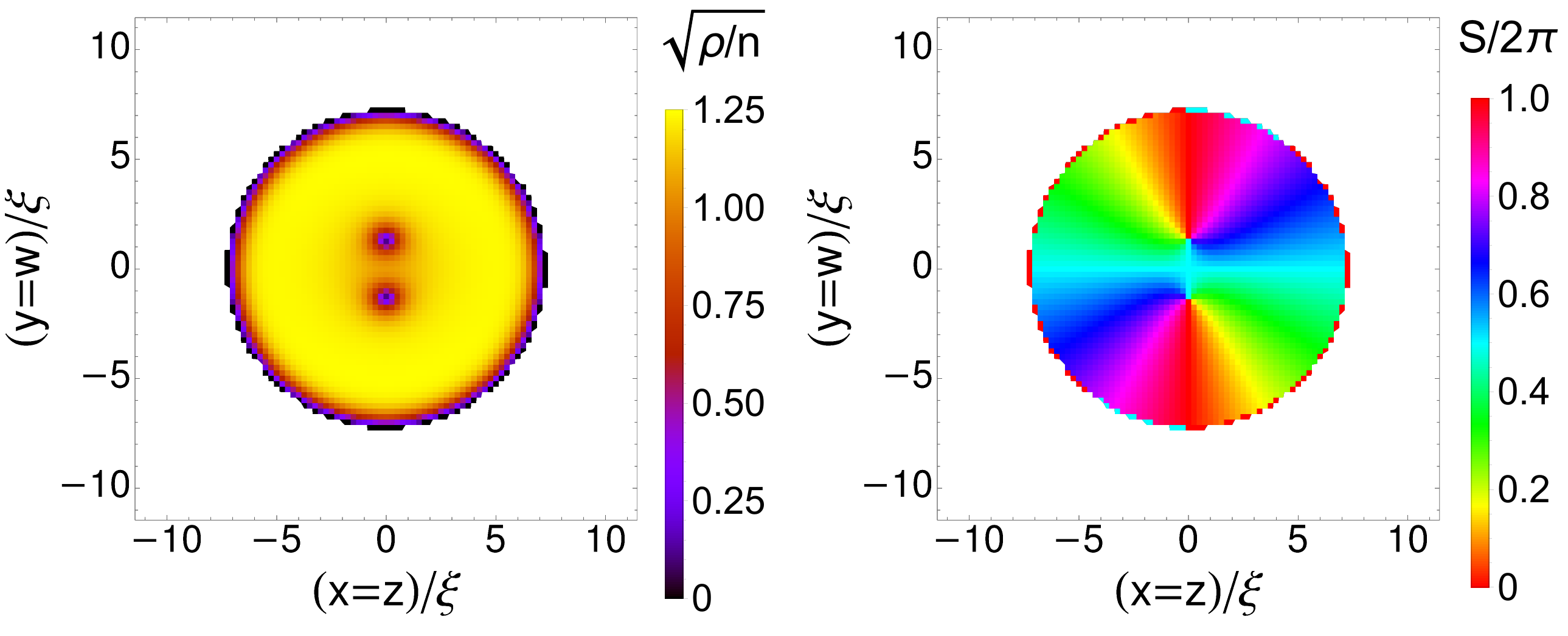}
    \end{minipage}
    \begin{minipage}{0.45\textwidth}
        \includegraphics[width=\linewidth]{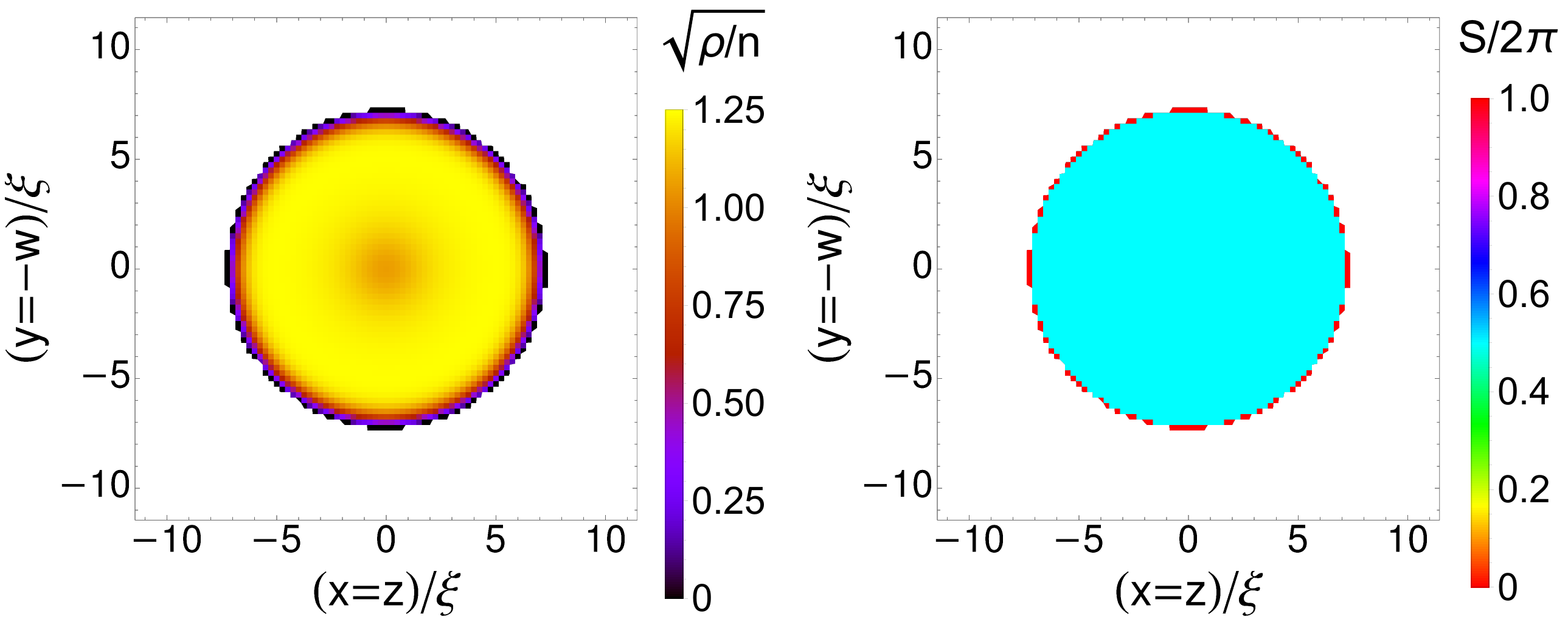}
    \end{minipage}
    \begin{minipage}{0.45\textwidth}
        \includegraphics[width=\linewidth]{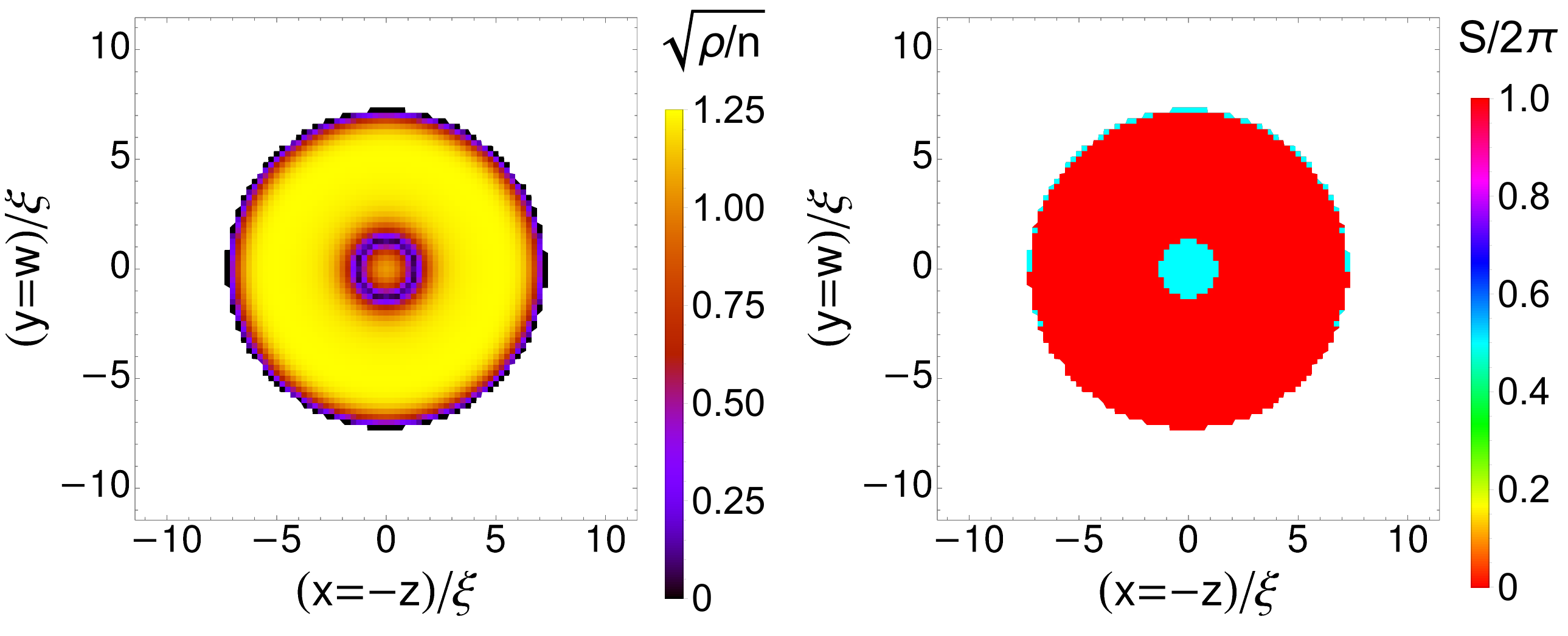}
    \end{minipage}
    \begin{minipage}{0.45\textwidth}
        \includegraphics[width=\linewidth]{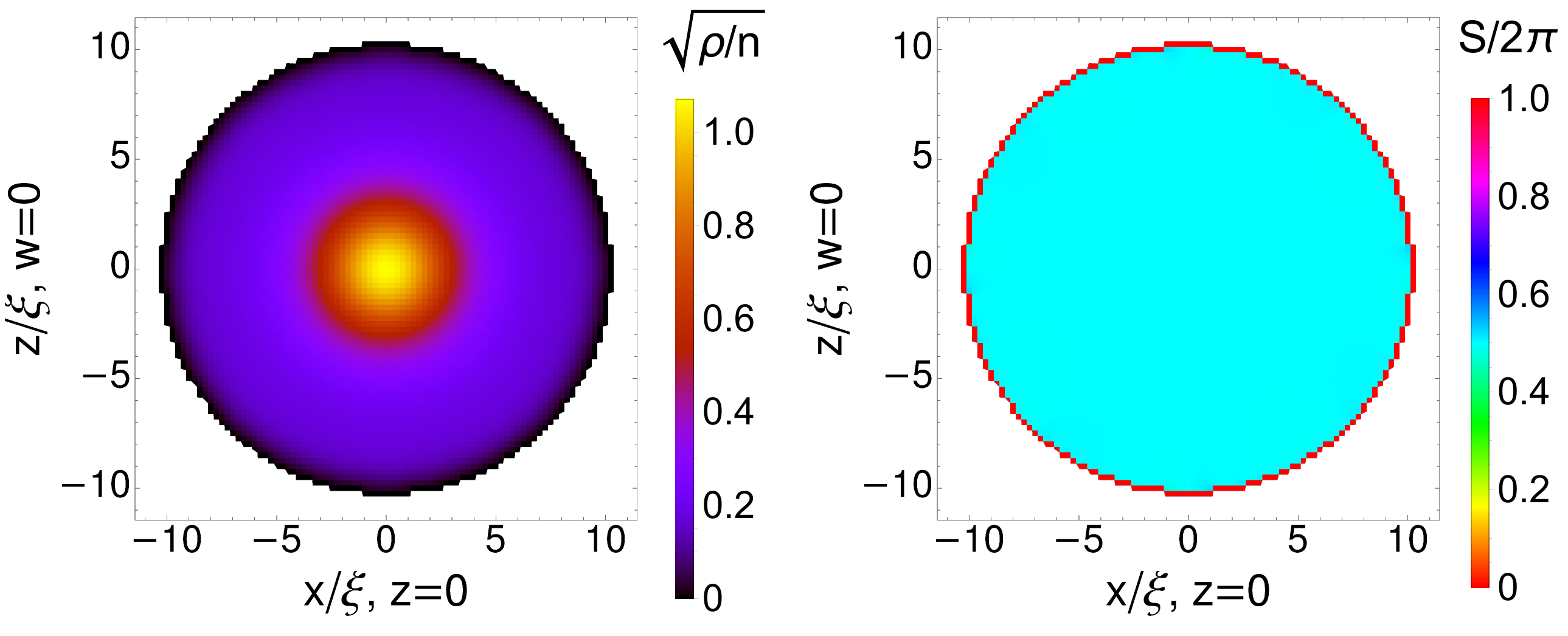}
    \end{minipage}
    \caption{2D cuts of the density and phase profile of the state whose vortex core is shown in Fig~\ref{fig:HiResF1.50AntiSkew}. The top and center rows show cuts that were previously used to visualise the orthogonal intersecting state in Ref~\cite{mccanna2021}, allowing us to compare these two states. The bottom row shows cuts that were not previously used, since, by symmetry,  they gave density and phas profiles that were seen before. Now these symmetries are broken and these cuts give useful information.}
    \label{fig:2DCuts}
\end{figure*}

\section{Numerical Methods}
\label{app:methods}

In this appendix, we will briefly review the numerical methods used in Sec.~\ref{sec:Equal}. As in our earlier works~\cite{mccanna2021, mccannaunequal}, we use the imaginary time evolution method (ITEM) to find solutions of the doubly-rotating 4D GPE [Eq~\eqref{eq:GPE4DR}], where we apply second-order finite differences in space and a first order discretisation in time. We perform all calculations using a Cartesian grid within a 4D hypersphere. This hypersphere has a radius set by \(N_{\text{grid}}\approx41\) gridpoints, with hardwall boundaries imposed on boundary points (i.e. those points with less than 8 nearest neighbours). In total, this corresponds to having a total number of gridpoints of approximately \(1.4\times10^{7}\). For most calculations, the spatial step size is given by \(\Delta x = 0.5\xi\), in order that we consider a large system of radius \(R\approx21\xi\), reducing the importance of boundary effects.

The expected critical frequency \(\Omega_c\) is calculated from Eq.~\eqref{eq:4Dcrit} with \(R\) set to \(N_{\text{grid}}\Delta x - \xi\), i.e. to approximately account for the boundary region, we subtract one healing length. Numerical results suggest that, for \(\Delta x=0.5\xi\), a more accurate value for the critical frequency of \(0.9\Omega_c\)~\cite{mccannaunequal}; this is likely due to both finite size effects and approximations in the theoretical derivation of the critical frequency.

To construct the initial states for the ITEM, we build up the order parameter from a suitable density profile and phase profile, adding noise (up to $20\%$ of the background value) to both the real and imaginary parts of \(\psi\). In keeping with our tilted plane ansatz in Sec.~\ref{sec:Equal}, the initial density profile is homogeneous except for at the boundary where it is smoothly goes to zero. The initial phase factor is set by the vortex configuration expected at low energy for the chosen parameters. We deem the ITEM to have converged once the relative variations in the particle number, \(N\) (calculated as the sum of \(\abs{\psi}^2\Delta x^4\)), and chemical potential (the sum of the LHS of Eq~\eqref{eq:GPE4DR} multiplied by \(\psi^*\Delta x^4/N\)), between iterations reaches below \(10^{-10}\). We then output the order parameter and calculate the corresponding energy~\cite{mccannaunequal}. We also output the coordinates of all points making up the vortex core, where a point is deemed to be in the core if \(|\psi|\) is less than the spatial resolution \(\Delta x/\xi\) and if the point is over a healing length away from the boundary. The former criterion is motivated by the fact that the order parameter vanishes linearly as one approaches a singly charged vortex core~\cite{pitaevskii2003}. We then plot the location of the vortex core by combining a 3D scatter plot (representing the \(x\), \(y\), and \(z\) coordinates) with colour (representing the \(w\) coordinate).

\section{Additional Numerical Results}
\label{app:numerical}

In this appendix, we will briefly present extra numerical results to supplement those in the main text. 
Fig~\ref{fig:HiResF1.50AntiSkew} shows the core plot for a final state with the same frequencies (\(\Omega=1.5\Omega_c\)) as Fig~\ref{fig:F1.50AntiSkew} but with a spatial step of \(\Delta x=0.25\xi\), corresponding to a smaller radius of \(R\approx10.3\xi\). As we can see in the third panel, this state also has no visible skewness. Also, this state exhibits the same kind of avoided crossing, as in Fig~\ref{fig:F1.50AntiSkew}, but with a smaller value for \(r_{min}\) as shown in  Fig~\ref{fig:AntiTiltDist}.


Fig~\ref{fig:2DCuts} shows the density and phase profiles of this same state along 2D cuts given by \(y=0 \ \& \ w=0\), \(x=y \ \& \ z=w\), \(x=z \ \& \ y=w\), \(x=z \ \& \ y=-w\). Note that a cut given by \(x=y \ \& \ z=w\) would show the same overall structure as that in the \(y=0 \ \& \ w=0\) cut, but in a smaller disc due to the cut being diagonal. These particular cuts were chosen as they were used in the appendix of our previous paper~\cite{mccanna2021} to visualise the orthogonal intersecting state. Comparing those previous plots to those in Fig~\ref{fig:2DCuts} gives us an idea of how the avoided crossing affects the phase as well as the density.

In more detail, firstly, the top left density plot in Fig~\ref{fig:2DCuts}  has gone from showing orthogonal intersecting lines of depletion in Ref.~\cite{mccanna2021} to what look hyperbolae, while the top left phase plot shows that the phase jumps across these lines in the exact same way as before. Secondly, the top right plots show a cut that has gone from orthogonal intersecting vortex lines to density depletions that do not quite reach zero and smoothly
vanish near the origin. This is because the vortex cores are curving out of the plane of the cut in this case, and the corresponding phase jumps become smooth variations. Thirdly, the center left plots show a cut that was previously essentially a plot of a doubly charged point vortex but now appears to be a pair of singly like-charged point vortices with a small separation. Fourthly, the center right cut was previously in Ref.~\cite{mccanna2021} a strange case where the density showed what looked like a point vortex, but because of the cut chosen the phase was constant. Now the perturbed version shows that this density depletion no longer goes to zero. Finally, the bottom row shows two more 2D cuts of this state which did not show any new features for the intersecting orthogonal state due to its symmetries. Since the avoided crossing state has fewer symmetries these plots are now interesting. The bottom left plots correspond to a cut (\(x=-z \ \& \ y=w\)) that previously had a zero in density but no phase winding, but now show a vortex ring, with a corresponding phase jump. The top right shows plots that are centred on the \(x\en y\) plane, and hence would have shown very low density (up to numerical accuracy) for the orthogonal state. Now, due to the avoided crossing, we see a region of nonzero density around the origin that reduces to zero as the radius increases, with the corresponding phase plot being constant. 

\bibliographystyle{apsrev4-2}

\end{document}